\documentclass[10pt]{iopart} 
\def\onlinecite{\cite}   % absent in IOP
\def\affiliation{\address}      % absent in IOP
\usepackage{bm,graphicx,iopams,setstack,color}
\begin{document}
\title[]{Floquet-Weyl and Floquet-topological-insulator phases in a stacked two-dimensional ring-network lattice}
\author{Tetsuyuki Ochiai}
\affiliation{Research Center for Functional Materials, National Institute for Materials Science (NIMS), Tsukuba 305-0044, Japan}
\date{\today}

\begin{abstract}
We show the presence of Floquet-Weyl and Floquet-topological-insulator phases in a stacked two-dimensional ring-network lattice. The Weyl points in the three-dimensional Brillouin zone and Fermi-arc surface states are clearly demonstrated in the quasienergy spectrum of the system in the Floquet-Weyl phase. In addition, chiral surface states coexist in this phase. The Floquet-topological-insulator phase is characterized by the winding number of two in the reflection matrices of the semi-infinite system and resulting two gapless surface states in the quasienergy gap of the bulk.  The phase diagram of the system is derived in the two-parameter space of hopping S-matrices among the rings. We also discuss a possible optical realization of the system together with the introduction of synthetic gauge fields. 
\end{abstract}
%\pacs{73.20.At,03.65.Vf,42.79.Gn}
%\keywords{}

(To appear in {\it J. Phys.: Condens. Matter})

\maketitle
\ioptwocol

\section{Introduction}

In the last decade, intense research activities in topological insulator (TI) and topological superconductor \cite{RevModPhys.83.1057} have induced the investigation of topological phenomena in other areas of research. 
Topological photonics \cite{lu2014topological}, acoustics \cite{PhysRevLett.114.114301}, magnonics \cite{PhysRevB.87.174427}, phononics \cite{PhysRevLett.115.104302}, are such categories.  In photonics, various two-dimensional (2D) topological phenomena have been predicted theoretically and confirmed experimentally \cite{haldane2008prd,wang2008rfo,yu2008owe,ochiai2009photonic,rechtsman2013photonic,khanikaev2013photonic,hafezi2013imaging,PhysRevLett.110.203904,ochiai2015time,PhysRevLett.114.223901,xu2016accidental}. They yield unidirectional flows of light, being robust against disorder. They also give a positive feedback to TI physics, because  different setups and viewpoints are provided by topological photonics.

In contrast to 2D topological phenomena, it is still embryonic to investigate three-dimensional (3D) topological phenomena in photons. So far, limited works such as photonic topological crystalline insulators \cite{PhysRevB.84.195126,lu2016symmetry} and photonic Dirac or Weyl semimetals \cite{lu2013weyl,lu2015experimental,PhysRevB.93.235155,chen2015experimental} have been reported.      
This is partly because they require sophisticated designs of complex 3D photonic architectures.

Recently, Wang et al. proposed a novel construction of 3D topological phases in a 3D waveguide network \cite{PhysRevB.93.144114}, having its optical realization in mind. The proposed system can be regarded as a Floquet-Bloch one, although no periodic drive is introduced. It has Floquet-Weyl (FW) and Floquet-topological-insulator (FTI) phases, hosting topologically gapless and gapped band structures, respectively.   If such a system is realized experimentally in optics, many interesting physics can be examined. Anderson localization, synthetic gauge fields, and possible interaction effects (via cavity quantum electrodynamics) in the FW and FTI phases can be investigated in an optical (bosonic) platform.

To accelerate this program, here, we present a simple alternative of the 3D waveguide network, namely, a stacked 2D ring network. It consists of 2D ring network layers stacked in the third direction with interlayer scattering channels.  
The 2D ring network has shown its potential for various 2D topological phenomena
 and can be realized in silicon photonics technologies \cite{hafezi2013imaging,PhysRevLett.110.203904,PhysRevB.89.075113}. 
As shown in this paper, the stacked 2D ring network has a similar topological phase structure as in the 3D waveguide network.  This construction seems to be simpler than in the waveguide network, and various interesting phenomena of the monolayer 2D ring network can be integrated.

The network-model description is physically more fruitful than the tight-binding-model description, which is often employed in various contexts. Moreover, the ring-network model has a high fidelity in corresponding ring-resonator lattices realized by silicon photonics. The network model includes a tight-binding-model picture in its weak-coupling regime \cite{poon2004matrix}, and can describe topological phenomena in a strong-coupling regime, which are not available in the tight-binding model. Therefore, the network model is a good starting point to study topological phases in photonics.

Our construction of the FW and FTI phases are reminiscent of the layered 2D TIs \cite{PhysRevLett.107.127205} and layered Chern insulators \cite{PhysRevLett.116.066401}. There, Weyl-semimetal and 3D TI phases emerge. In our system, the quasienergy winding inherent in Floquet systems plays a crucial role as in the 3D waveguide network. Moreover, our system is completely bosonic, having its optical realization in mind. These properties show a striking contrast to the layered electronic systems.

Related 3D network models have been investigated so far, mainly focusing on the Anderson localization in layered 2D quantum-Hall and quantum-spin-Hall systems \cite{PhysRevLett.75.4496,klesse1999modeling,PhysRevB.89.155315}. A different design of 3D network models was introduced by the present author, toward a photonic counterpart of topological crystalline insulator with time-reversal symmetry \cite{ochiai2015gapless}.     
Here, we deal with a system without disorder and time-reversal symmetry and focus on its spectrum as a Floquet-Bloch system.

Our system can be viewed as a Floquet system without a periodic drive. Conventional Floquet systems are laser-driven ones, in which nontrivial topology can emerge by irradiating electronic systems with a continuous-wave light \cite{PhysRevB.79.081406,PhysRevLett.105.017401,lindner2011floquet,wang2013observation,PhysRevB.89.121401}.   
There, the problem is reduced formally to the diagonalization of the time-translation unitary operator for one period. However, it is often the case that the operator is not easily obtained. In our case, the time-translation operator is replaced by a certain unitary matrix to be diagonalized. 
The matrix is analytically available in a simple form. A similar setting as ours takes place in the discrete-time quantum walk, where various topological phases can emerge \cite{PhysRevB.86.195414,kitagawa2012observation,PhysRevB.92.045424}.

In this paper, we show theoretically that the stacked 2D ring network layers exhibits both the FW and FTI phases. A detailed analysis on the phase structure is given. The analysis indicates that the system has robust FW phases bounded by line-node phases. Through a collapse of Weyl points into line nodes, the system can exhibit FTI phases with gapless surface states. A topological characterization of the FTI phases is made, based on the winding numbers of relevant S-matrices.

This paper is organized as follows. In Sec. II, we present our model and analytically solve the eigenvalue equation of the Bloch modes in the bulk. By studying the possible degeneracy of the eigenvalues, the phase diagram is obtained regarding the gap.  Section III is devoted to present the surface states using the S-matrices of slab systems. Topological characterization of the phase diagram is also made. In Sec. IV, an optical realization of the network model and an implementation of synthetic gauge fields are discussed. Finally in Sec. V, we summarize the results.

\section{Model}

\begin{figure}
\begin{center}
\includegraphics[width=0.45\textwidth]{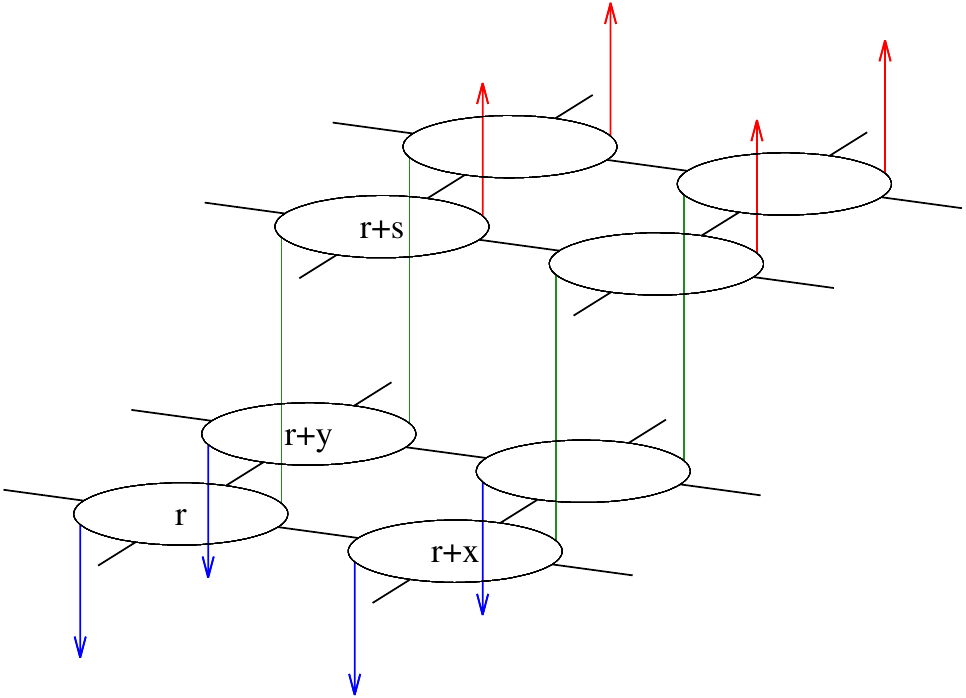} 
\includegraphics[width=0.23\textwidth]{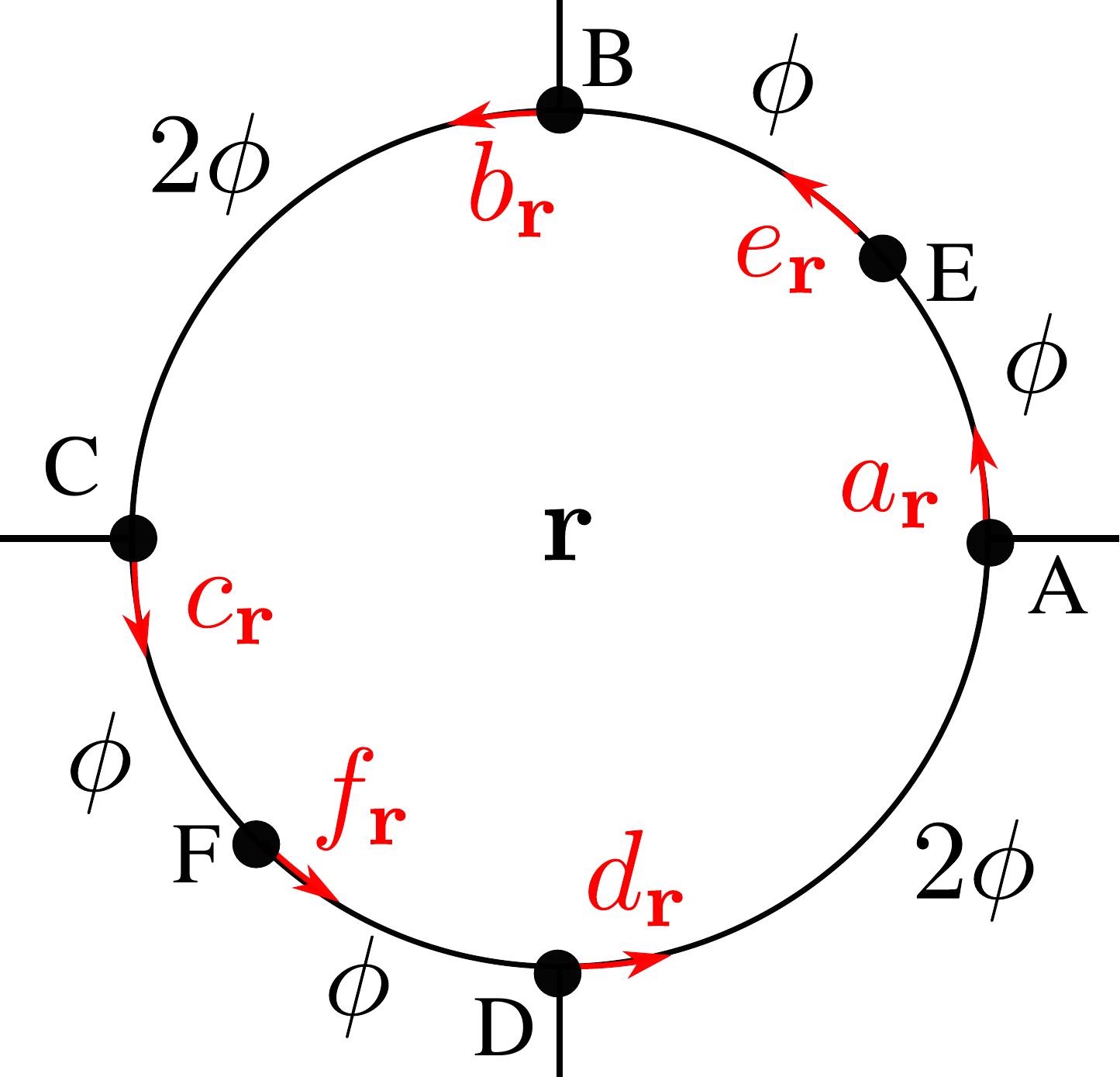}
\end{center} 
\caption{\label{Fig_model}(Color online) The stacked 2D ring network model under study. It consists of the 2D square-lattice ring-network layers stacked in the third direction with the in-plane shift.  The shift vector ${\bm s}$ is (1/2,1/2,1/2) in units of the lattice constant. Solely the counter-clockwise flow of the ring propagation mode is considered. Mode amplitudes $\alpha_{\bm r}$ ($\alpha=a,b,c,d,e,f$) are defined at the nodes A,B,C,D,E, and F.  The nodes A,B,C, and D connect to nearby rings in plane. The node E connects to the node F of the upper nearby ring. The propagation phase of the mode for every 1/8 propagation of the ring is denoted as $\phi$.}
\end{figure}

Let us first present our model of the stacked 2D ring network. It consists of a multilayer stacking of the 2D ring network.  Each layer is arranged in the square lattice as shown in Fig. \ref{Fig_model}.  
%%% Fig. 1 %%%%%%%%%%%%%%%%%%%%%%
%%%%%%%%%%%%%%%%%%%%%%%%%%%%%%%%%%%%%%%%
The layer-by-layer stacking is made in such a way that the adjacent layer has the lateral shift of (1/2,1/2) and vertical shift of 1/2 in units of the lattice spacing. The rings are connected in plane at nodes A,B,C, and D, and upward and downward at nodes E and F, respectively. In each ring, we introduce a counter-clockwise propagating mode whose propagation phase is denoted as $\phi$ for every 1/8 propagation in the rings \footnote{The clockwise propagating mode is assumed to decouple from the counter-clockwise one. See the discussion in Sec. 4.}.  This phase is self-consistently determined in the periodic system  where the Bloch theorem is applied. The mode is scattered at the nodes. Hence, we introduce the mode amplitudes $\alpha_{\bm r}$ at each node.

The scattering at the nodes is described by the hopping S-matrices with 
\begin{eqnarray}
&\left(\begin{array}{c}
a_{\bm r}\\
c_{{\bm r}+\hat{x}}
\end{array}\right)=S_1 \left(\begin{array}{c}
d_{\bm r}\\
b_{{\bm r}+\hat{x}}
\end{array}\right){\rm e}^{2{\rm i}\phi},\\
&\left(\begin{array}{c}
b_{\bm r}\\
d_{{\bm r}+\hat{y}}
\end{array}\right)=S_2 \left(\begin{array}{c}
e_{\bm r}\\
f_{{\bm r}+\hat{y}}
\end{array}\right){\rm e}^{{\rm i}\phi},\\
&\left(\begin{array}{c}
e_{\bm r}\\
f_{{\bm r}+{\bm s}}
\end{array}\right)=S_3 \left(\begin{array}{c}
a_{\bm r}\\
c_{{\bm r}+{\bm s}}
\end{array}\right){\rm e}^{{\rm i}\phi},
\end{eqnarray}
where ${\bm s}=(\hat{x}+\hat{y}+\hat{z})/2$ being $\hat{\mu}$ ($\mu=x,y,z$) the unit vector pointing $+\mu$ direction.  
We will discuss in Sec. IV, how such a scattering can be realized in optical systems.

The scattering is inversion symmetric, because in each scattering element, there is the inversion symmetry with respect to the center of two adjacent nodes. 
The S-matrices thus satisfy 
\begin{eqnarray}
S_i=\sigma_1S_i\sigma_1 \quad (i=1,2,3), \label{Eq_invsym}
\end{eqnarray}
being $\sigma_1$ the Pauli matrix. Therefore, they are written by 
\begin{eqnarray}
&S_1=S_2={\rm e}^{{\rm i}\alpha}\left(\begin{array}{cc}
\cos\beta  & {\rm i}\sin\beta\\
{\rm i}\sin\beta & \cos\beta\end{array}
\right), \label{Eq_beta}\\ 
&S_3={\rm e}^{{\rm i}\gamma}\left(\begin{array}{cc}
\cos\delta  & {\rm i}\sin\delta\\
{\rm i}\sin\delta & \cos\delta\end{array}
\right).\label{Eq_delta}
\end{eqnarray}

Assuming the perfect 3D periodicity of the structure, the eigenvalue equation for the bulk modes is derived with the aid of the Bloch theorem. It becomes a Floquet-Bloch form, namely, a diagonalization of the unitary matrix $U_{\bm k}$ as 
\begin{eqnarray}
&U_{\bm k}\left(\begin{array}{c}
a_{\bm r}\\
c_{\bm r}\end{array}\right)={\rm e}^{-4{\rm i}\phi}\left(\begin{array}{c}
a_{\bm r}\\
c_{\bm r}\end{array}\right),
\end{eqnarray}
where 
\begin{eqnarray}
&U_{\bm k}=\tilde{S}_1\tilde{S}_2\tilde{S}_3,\\
&\tilde{S}_1=\left(\begin{array}{cc}
1 & 0\\
0 & {\rm e}^{-{\rm i}k_x}
\end{array}\right)S_1 \left(\begin{array}{cc}
0 & 1\\
{\rm e}^{{\rm i}k_x}& 0
\end{array}\right), \\ 
&\tilde{S}_2=\left(\begin{array}{cc}
1 & 0\\
0 & {\rm e}^{-{\rm i}k_y}
\end{array}\right)S_2\left(\begin{array}{cc}
1 & 0\\
0 & {\rm e}^{{\rm i}k_y}
\end{array}\right),\\
&\tilde{S}_3=\left(\begin{array}{cc}
1 & 0\\
0 & {\rm e}^{-{\rm i}k_3}
\end{array}\right)S_3\left(\begin{array}{cc}
1 & 0\\
0 & {\rm e}^{{\rm i}k_3}
\end{array}\right),\\
&k_3={\bm k}\cdot{\bm s}=\frac{1}{2}(k_x+k_y+k_z). 
\end{eqnarray}
Here, the propagation phase $\phi$ acts as the quasienergy in the Floquet system.

The system has the space-inversion symmetry as 
\begin{eqnarray}
U_{-{\bm k}}=\sigma_1 U_{\bm k} \sigma_1,
\end{eqnarray}
coming from the inversion symmetry, Eq. (\ref{Eq_invsym}) of the S-matrices. 
Therefore, the quasienergy spectrum satisfies $\phi_{-{\bm k}}=\phi_{\bm k}$. 
The time-reversal symmetry is broken, because solely the counter-clockwise mode is considered. The broken time-reversal symmetry can be implemented in optical ring-resonator systems as explained in Sec. 4.

We can analytically solve the eigenvalue equation. In doing so, we put $\alpha=\gamma=0$, because  these parameters merely shift the quasienergy in the bulk band structure. 
The two eigenvalues are  given by 
\begin{eqnarray}
\phi&=-\frac{\theta}{4},\quad \frac{\pi}{4} + \frac{\theta}{4}\quad \left({\rm mod}\; \frac{\pi}{2}\right),\\ 
\theta&=\sin^{-1}\xi, \\
\xi&=\sin\beta\cos\beta\cos\delta (\cos k_x + \cos k_y) \nonumber \\
&+\cos^2\beta\sin\delta \cos\left(\frac{k_x+k_y+k_z}{2}\right) \nonumber \\
&-\sin^2\beta\sin\delta \cos\left(\frac{k_x+k_y-k_z}{2}\right). 
\end{eqnarray}
Possible degeneracy in the quasienergy spectrum takes place only at $\phi=\pm \pi/8$.

The degeneracy is, most possibly, point-like. Namely, the degeneracy takes place at certain points in the 3D Brillouin zone. 
Let us study in detail how this happens.  The degenerate point corresponds to $\xi=\pm 1$, the minimum and maximum values of $\xi$ under real $\beta,\delta$, and $k_\mu$'s.  Since $\xi$ is invariant under the exchange of $k_x$ and $k_y$, the minimum and maximum should appear on the plane of $k_x=k_y$ in the Brillouin zone.   The minimum and maximum require 
$\partial \xi/\partial k_\mu =0\quad (\mu=x,z)$ on this plane,  so that the degeneracy takes place most probably at certain points there.

Moreover, this point-like degeneracy takes place simultaneously at four points in the Brillouin zone. This is because $\xi$ satisfies the following relation: 
\begin{eqnarray}
\xi(\pi-k_x,\pi-k_y,-k_z)=-\xi(k_x,k_y,k_z).
\end{eqnarray}  
Together with the inversion symmetry, the degeneracy takes place simultaneously at 
\begin{eqnarray}
{\bm k}={\bm k}^c,-{\bm k}^c,(\pi-k_x^c,\pi-k_y^c,-k_z^c),(k_x^c-\pi,k_y^c-\pi,k_z^c)
\end{eqnarray}

We can show that the point-like degeneracy is of the Weyl type. To confirm this fact, let us consider the derivative of $\theta$ with respect to ${\bm k}$ around the degenerate point ${\bm k}^c$. It becomes 
\begin{eqnarray}
\frac{\partial \theta}{\partial k_\mu}=
\frac{\partial \xi}{\partial k_\mu}\frac{1}{\sqrt{1-\xi^2}}. 
\end{eqnarray}
Around the degenerate point, $\xi$ can be expanded as 
\begin{eqnarray}
\xi=\pm\left(1-\sum_{\mu,\nu} C_{\mu\nu}(k_\mu-k_\mu^c)(k_\nu-k_\nu^c)\right). 
\end{eqnarray}
Therefore, the derivative of $\theta$ is expressed as 
\begin{eqnarray}
\frac{\partial \theta}{\partial k_\mu}\simeq \frac{\displaystyle{ \mp 2\sum_{\nu}C_{\mu\nu}(k_\nu-k_\nu^c)}}{\sqrt{\displaystyle{2\sum_{\nu,\rho} C_{\nu\rho}(k_\nu-k_\nu^c)(k_\rho-k_\rho^c) }}},   
\end{eqnarray}
which corresponds to a Weyl-type dispersion with the spatial anisotropy.

\begin{figure}
\centerline{
\includegraphics[width=0.225\textwidth]{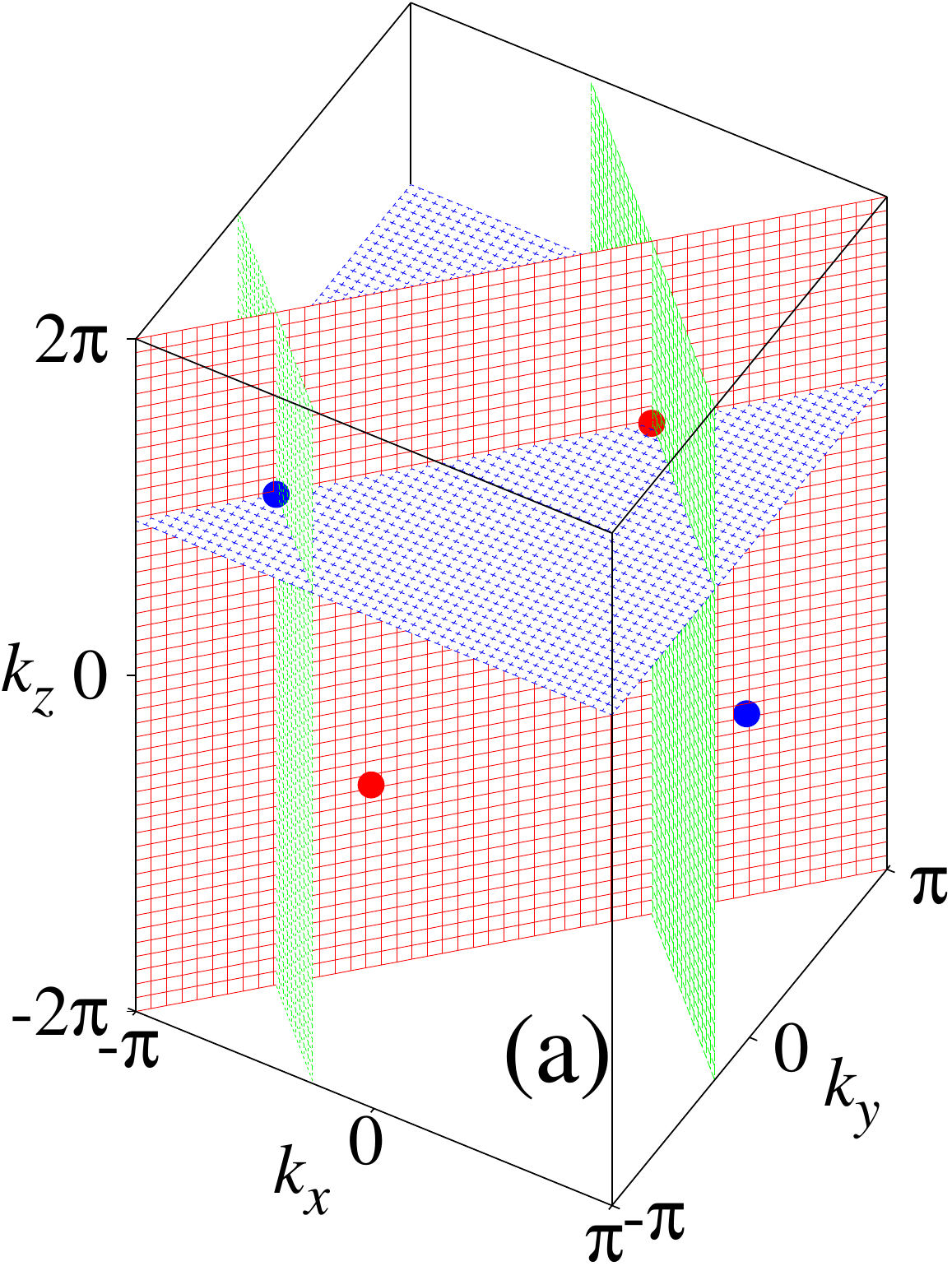} 
\includegraphics[width=0.225\textwidth]{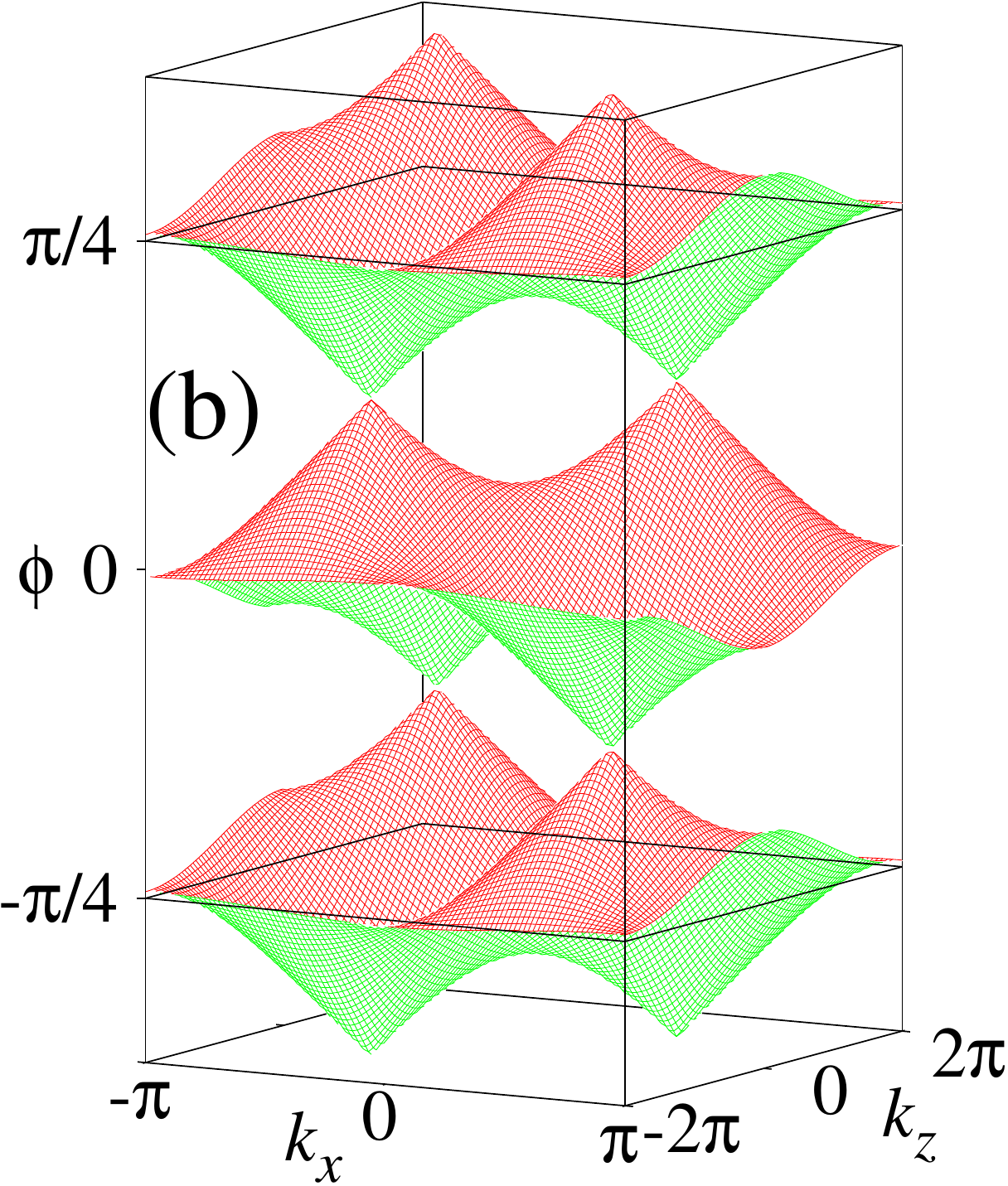} }
\centerline{
\includegraphics[width=0.225\textwidth]{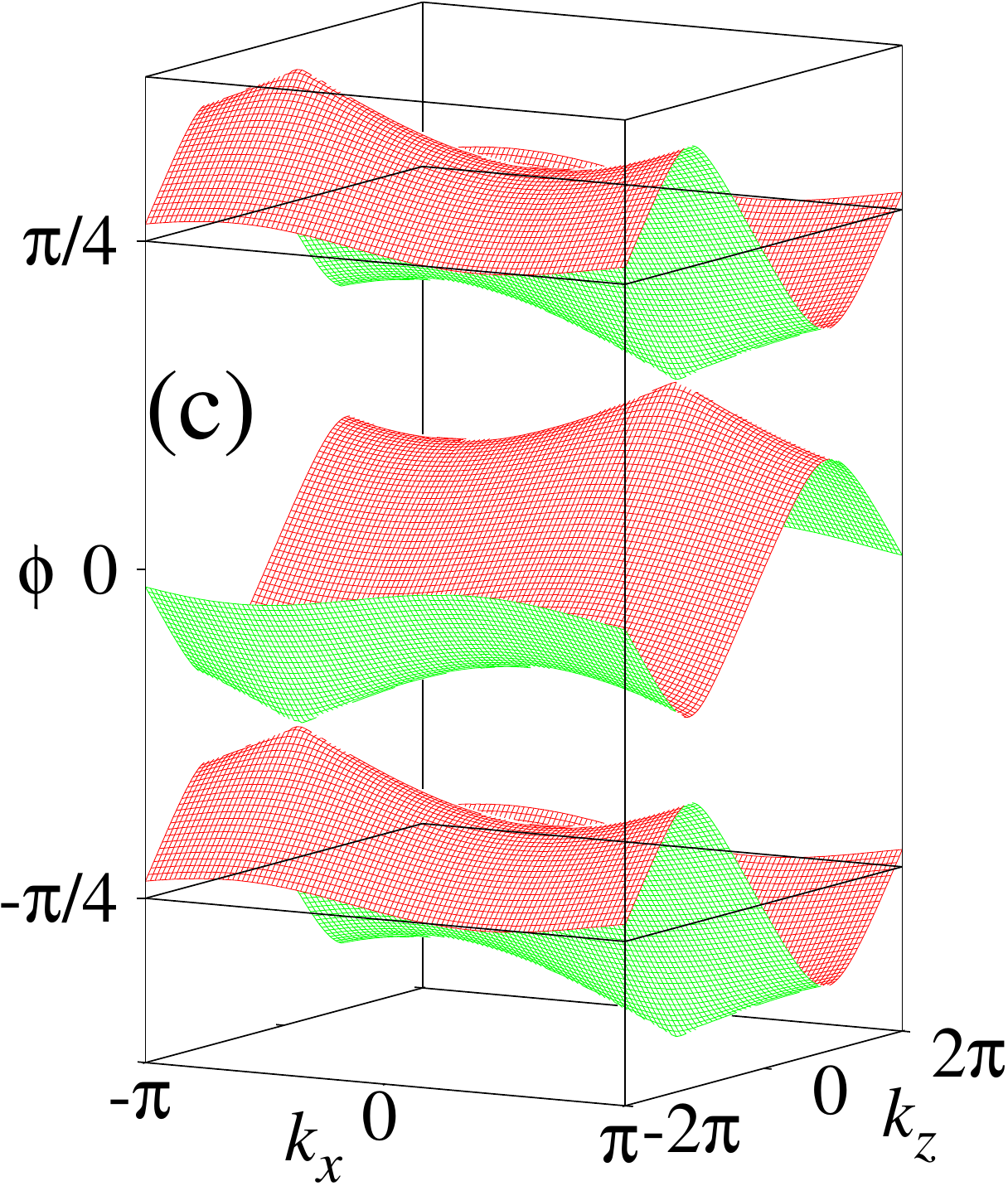} 
\includegraphics[width=0.225\textwidth]{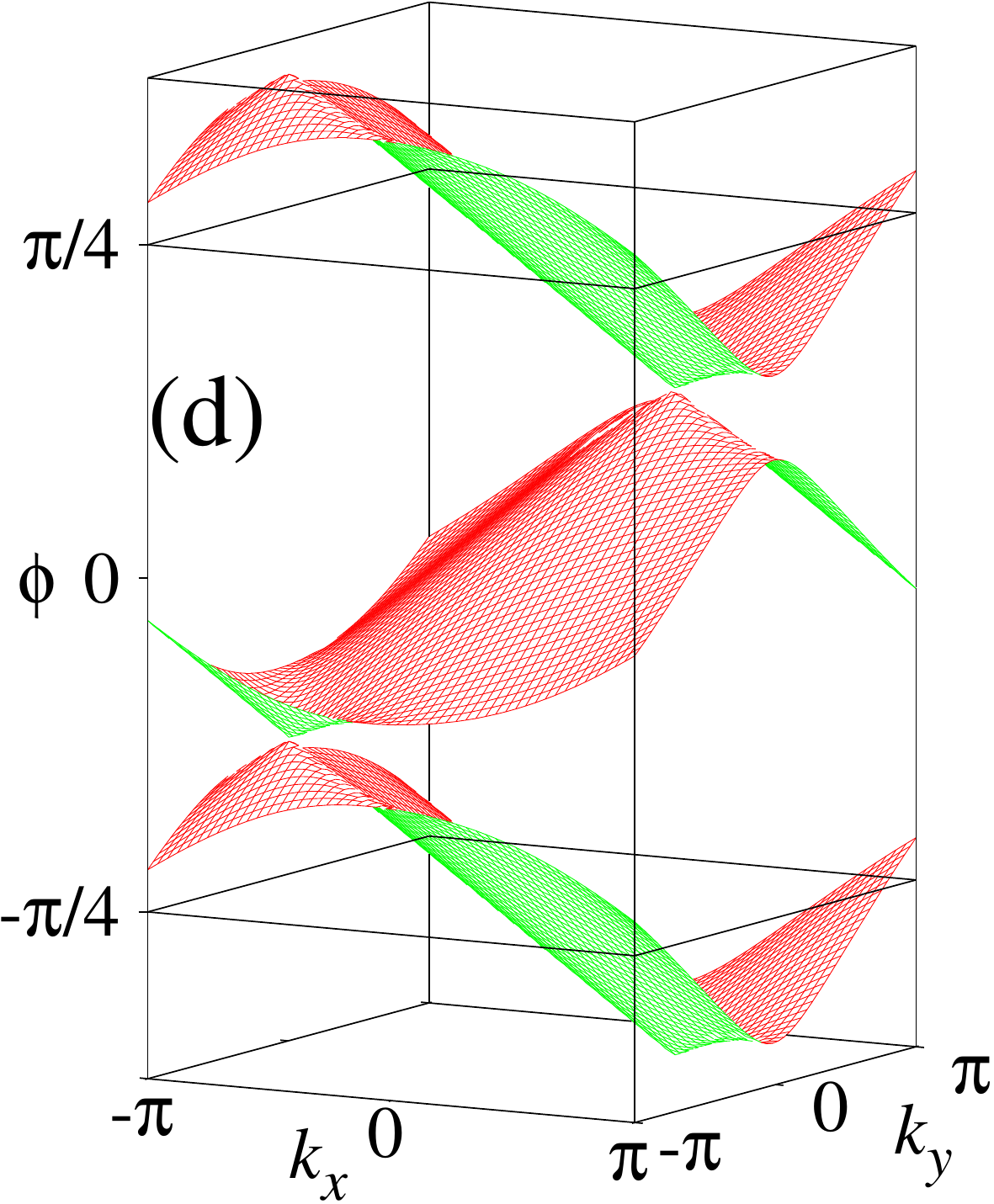} }
%\centerline{
%\includegraphics[width=0.45\textwidth]{fig2a.pdf} 
%\includegraphics[width=0.45\textwidth]{fig2b.pdf} }
%\centerline{
%\includegraphics[width=0.45\textwidth]{fig2c.pdf} 
%\includegraphics[width=0.45\textwidth]{fig2d.pdf} }
\caption{\label{Fig_weyl} (Color online) Position of the Weyl points and the quasienergy band structure at $\beta/(2\pi)=0.15$ and $\delta/(2\pi)=0.31$. (a) Four Weyl points in the Brillouin zone are denoted by solid dots. The inversion symmetric pair is denoted by the same color.  The quasienergy band structures on the $k_x=k_y$ plane (b), on the   $k_y=-k_x+2k_x^c$  plane (c), and  on the $k_z=k_z^c>0$ plane (d). These planes are depicted in (a). }
\end{figure}

Figure \ref{Fig_weyl} shows a typical band structure of the quasienergy with four Weyl points. 
%%%%  Fig. 2 %%%%%%%%%%%%%%%%%%%%%%%%%%%%%%%
%%%%%%%%%%%%%%%%%%%%%%%%%%%%%%%%%%%%%%%%%%%%%5
The Weyl points appear at $\phi=\pm \pi/8$ as expected. Owing to the inversion symmetry, they also appear in the pair of ${\bm k}=\pm {\bm k}_c$.   We can see clearly the linear dispersion around the Weyl points.

Exceptionally, under a certain condition for $\beta$ and $\delta$, the degeneracy of line- and surface-node types takes place. For instance, at generic points on $\delta=\pm\pi/2$, line nodes emerge at $k_x+k_y=\pm \pi$ on the $k_z=\pm\pi$ plane. These line nodes become a surface node at particular points of $\beta$. At $\beta=0$ and $\pm\pi$, the nodal surfaces are given by $k_x+k_y+k_z=2n\pi$ ($n$: integer). At $\beta=\pm\pi/2$, the nodal surfaces are $k_x+k_y-k_z=2n\pi$.

\begin{figure}
\begin{center}
\includegraphics[width=0.45\textwidth]{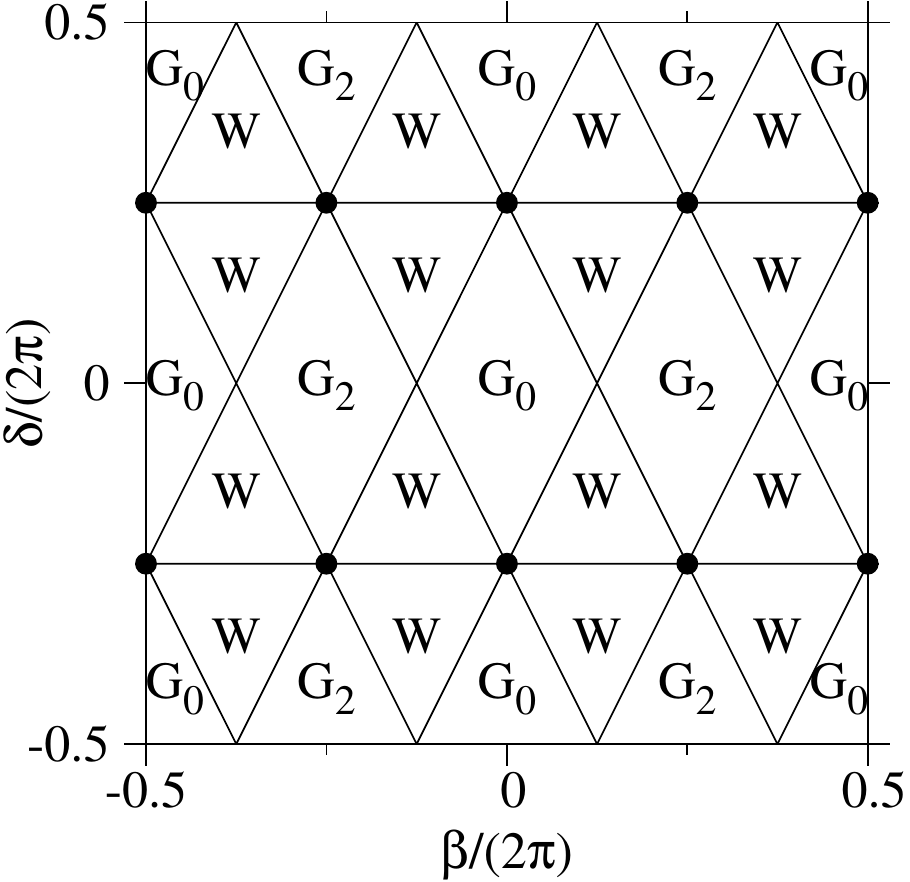} 
\end{center}
\caption{\label{Fig_phase} Phase diagram regarding the gap in the bulk band structure of the network model. The phase space is spanned by two parameters, $\beta$ and $\delta$, defined in Eqs. (\ref{Eq_beta}) and (\ref{Eq_delta}). Symbols ``G'' and ``W'' represent the gapped and Weyl  phases, respectively. The solid lines that divide the two phases represent the line-node phase. Solid circles represent the surface-node phase. The gapped phases are further classified into two categories, nontopological and topological. The former and latter are denoted by ${\rm G}_0$ and ${\rm G_2}$, respectively, where the subscript represents the winding number $n$ defined in Eqs. (\ref{Eq_winding1}) and  (\ref{Eq_winding2}).}
\end{figure}

The phase diagram regarding the degeneracy in the bulk is plotted in Fig. \ref{Fig_phase}.
%%%%% Fig. 3 %%%%%%%%%%%%%%%%%%%%%%%%%%%%%%%%% 
%%%%%%%%%%%%%%%%%%%%%%%%%%%%%%%%%%%%%%%%%%%%%%%%%
Here, we have robust Weyl phases and gapped phases bounded by the line-node phases that are solely extended on lines in the phase diagram. It is important that the Weyl phases are not directly touching with the gapped phases. This property implies that the  Weyl points always disappear through the formation of a line node. It is not a simple pair annihilation of two Weyl points.

Finally, we remark that at $\delta=0$, the system becomes disconnected 2D ring network layers. There, we have the gap closing at $\beta=\pm \pi/4$ and $\pm 3\pi/4$ and the  anomalous Floquet insulator phase emerge in $\pi/4<|\beta|<3\pi/4$ \cite{PhysRevLett.110.203904}. These results are fully consistent with the phase diagram.

\section{Surface states}
It is well known that the Weyl point is accompanied by the  Fermi-arc surface states whose dispersion curve connects two Weyl points projected onto the surface Brillouin zone \cite{PhysRevB.83.205101}. 
In addition to the Fermi arc, our system can have gapless surface states whose origin is the chiral edge state of the 2D ring network \cite{PhysRevLett.110.203904}.  The 2D ring network is described by the Chalker-Coddington network model \cite{0022-3719-21-14-008} without disorder. Since the latter model was originally introduced to simulate quantum Hall systems, the 2D ring network  has chiral edge states analogous to those in quantum Hall systems \footnote{Strictly speaking, the chiral edge state in the 2D ring network does not simply come from a nonzero Chern number. Rather, it is due to a scenario inherent in the Floquet system like that given in Ref. \onlinecite{PhysRevB.82.235114}.
}. 
These two kinds of surface states make the system a bit complicated.

To explore the surface states, it is important to have the S-matrix for a slab system of the ring network.  With the S-matrix along with a boundary condition at the both surfaces of the slab, we can evaluate the surface-state dispersion relation. Moreover, topological properties of the bulk can be extracted from the slab S-matrix. By the so-called bulk-edge correspondence \cite{hatsugai1993cna}, gapless surface states that are robust against surface decorations are predicted from the S-matrix.

Let us first consider the slab in the $z$ direction. 
Suppose that the slab has  $N+1$ layers thickness in this direction. 
We introduce the slab S-matrix  as 
\begin{eqnarray}
&\left(\begin{array}{c}
f_N\\
e_0
\end{array}\right) =S_{z;N}
\left(\begin{array}{c}
a_0\\
c_N
\end{array}\right),\\
&{\alpha}_n=\alpha_{{\bm r+n{\bm s}}}\quad (\alpha=a,b,c,d,e,f).
\end{eqnarray} 
This S-matrix is constructed recursively as follows.
At $N=1$, the S-matrix is simply written as 
\begin{eqnarray}
S_{z;1}=\sigma_1 S_3{\rm e}^{{\rm i}\phi}.  
\end{eqnarray}
For the slab of thickness $N$, we divide it into upper and lower parts. 
Provided that the S-matrices of the upper and lower parts are given, then we have 
\begin{eqnarray}
&\left(\begin{array}{c}
f_N\\
e_i
\end{array}\right) =S_{z;u}
\left(\begin{array}{c}
a_i\\
c_N
\end{array}\right), \\
&\left(\begin{array}{c}
f_i\\
e_0
\end{array}\right) =S_{z;l}
\left(\begin{array}{c}
a_0\\
c_i
\end{array}\right). 
\end{eqnarray} 
On the other hand, we also know 
\begin{eqnarray}
&\left(\begin{array}{c}
a_i\\
c_i
\end{array}\right) =S_{\rm int}
\left(\begin{array}{c}
f_i\\
e_i
\end{array}\right),\\
&S_{\rm int}= {\rm e}^{3{\rm i}\phi}\tilde{S}_1\tilde{S}_2\sigma_1
\end{eqnarray}
By removing $\alpha_i$ ($\alpha=a,c,e,f$), we obtain 
\begin{eqnarray}
& S_{z;N}= \left(\begin{array}{ll}
S_{z;u}^{++}M^{++} & S_{z;u}^{++}M^{+-} + S_{z;u}^{+-}\\
S_{z;l}^{--}M^{-+} + S_{z;l}^{-+} & S_{z;l}^{--}M^{--}
\end{array}\right),\\ 
& M=A^{-1}B, \\
& A=\left(\begin{array}{cc}
1-S_{\rm int}^{+-} S_{z;u}^{-+} & 
 -S_{\rm int}^{++} S_{z;l}^{+-} \\
 -S_{\rm int}^{--} S_{z;u}^{-+} & 
1-S_{\rm int}^{-+} S_{z;l}^{+-}
\end{array}\right),\\
& B=\left(\begin{array}{cc}
S_{\rm int}^{++} S_{z;l}^{++} & 
S_{\rm int}^{+-} S_{z;u}^{--} \\
S_{\rm int}^{-+} S_{z;l}^{++} & 
S_{\rm int}^{--} S_{z;u}^{--}
\end{array}\right).
\end{eqnarray} 
Therefore, by starting from $N=1$, we can construct the slab S-matrix of arbitrary $N$, recursively.

As for the slab system having finite thickness in the $x$ direction, we introduce the slab S-matrix as 
\begin{eqnarray}
&\left(\begin{array}{c}
f_N\\
c'_N\\
b_0\\
b'_0
\end{array}\right) =S_{x;N}
\left(\begin{array}{c}
f_0\\
c'_0\\
b_N\\
b'_N
\end{array}\right), \label{Eq_slabx}\\
&\alpha_n=\alpha_{{\bm r}+n\hat{x}}, \quad  \alpha'_n=\alpha_{{\bm r}+{\bm s}+n\hat{x}}.
\end{eqnarray} 
The matrix $S_{x;1}$ is obtained from $S_\mu$ after lengthy calculation. 
The result is given in Appendix. 
The layer-doubling method, which is familiar in the field of low-energy electron-diffraction theory \cite{Pendry-LEED-book}, allows us to obtain $S_{x;N}$ from $S_{x;1}$.

In a pseudo gap, the transmission matrices (the Block diagonal parts of the slab S-matrix) vanish for sufficiently thick slab systems, because there is no bulk eigenmode to be excited. Therefore, the S-matrix is written as  
\begin{eqnarray}
S_{\mu;N}\to \left(\begin{array}{cc}
0 & R_\mu^{(u)}\\
R_\mu^{(l)} & 0\end{array}\right) 
\end{eqnarray}
with the reflection matrices $R_{\mu}^{(u)}$ and $R_{\mu}^{(l)}$ of the upper and lower surfaces, respectively. These matrices must be unitary.

Besides, the boundary condition for the eigenmodes in the slab system is generally written as 
\begin{eqnarray}
&a_0=Q_z^{(l)}e_0,\\
&c_N=Q_z^{(u)}f_N,
\end{eqnarray}
for the $z$ direction, and 
\begin{eqnarray}
&\left(\begin{array}{c}
 f_0\\
 c'_0 
\end{array}\right)=Q_x^{(l)}
\left(\begin{array}{c}
 b_0\\
 b'_0 
\end{array}\right), \\
&\left(\begin{array}{c}
 b_N\\
 b'_N 
\end{array}\right)=Q_x^{(u)}
\left(\begin{array}{c}
 f_N\\
 c'_N 
\end{array}\right)
\end{eqnarray}
for the $x$ direction. 
Here, $Q_\mu^{(u)}$ and $Q_\mu^{(l)}$ are unitary.   
For instance, if we introduced additional phase $\varphi_0$ at node F of the boundary rings (see Fig. \ref{Fig_model}), the boundary condition at the lower (bottom) surface of the slab in the $z$ direction is given by 
\begin{eqnarray}
f_0={\rm e}^{{\rm i}(\phi+\varphi_0)}c_0. \label{Eq_slabz_bcphase}
\end{eqnarray}
This boundary condition corresponds to  
\begin{eqnarray}
Q_{z}^{(l)}=S_{\rm int}^{+-} + S_{\rm int}^{++}{\rm e}^{{\rm i}(\phi+\varphi_0)}(1-{\rm e}^{{\rm i}(\phi+\varphi_0)}S_{\rm int}^{-+})^{-1}S_{\rm int}^{--}.
\end{eqnarray} 
As for the slab in the $x$ direction, we introduce the boundary condition as 
\begin{eqnarray}
&c_0 = {\rm e}^{{\rm i}(2\phi+p_0)}b_0, \label{Eq_slabx_bcphase}\\
&c'_0 = {\rm e}^{{\rm i}(2\phi+p'_0)}b'_0, \label{Eq_slabx_bcphase2}\\
&f_0 = {\rm e}^{{\rm i}(\phi+q_0)}c_0,\label{Eq_slabx_bcphase3}
\end{eqnarray}
by introducing additional phases at nodes C and F. 
This boundary condition results in 
\begin{eqnarray}
Q_{x}^{(l)}=\left(\begin{array}{cc}
{\rm e}^{{\rm i}(3\phi+p_0+q_0)} & 0 \\
0 & {\rm e}^{{\rm i}(2\phi+p'_0)}
\end{array}\right).  
\end{eqnarray}

The secular equation for possible surface states is thus given by 
\begin{eqnarray}
{\rm det}(1-R_\mu^{(l)}Q_\mu^{(l)})=0,  
\end{eqnarray}
for the lower surface. The equation for the upper surface states is obtained just by replacing superscript $(l)$ with $(u)$. Here, we concentrate on the lower surface.

In order to have a surface-state solution regardless of the boundary condition, 
$\Im[\log({\rm det}R_\mu^{(l)})]$ must cover the entire $(-\pi,\pi)$ as a function of momentum in the surface Brillouin zone \cite{PhysRevB.85.165409,PhysRevB.89.075113,poshakinskiy2015phase,PhysRevB.93.075405}. Here, $\Im$ stands for the imaginary part. 
In other words, it must have a nonzero winding number 
\begin{eqnarray}
n_{\mu\nu}^{(l)}=\int_{-\pi}^\pi \frac{{\rm d}k_\nu}{2\pi} \frac{\partial \Im[\log({\rm det}R_\mu^{(l)})]}{\partial k_\nu}, 
\end{eqnarray}  
where $k_\nu$ is the surface momentum in the $\nu$ direction. 
This $n_{\mu\nu}^{(l)}$ topologically classifies the gapped phases.

\begin{figure}
\begin{center}
\includegraphics[width=0.5\textwidth]{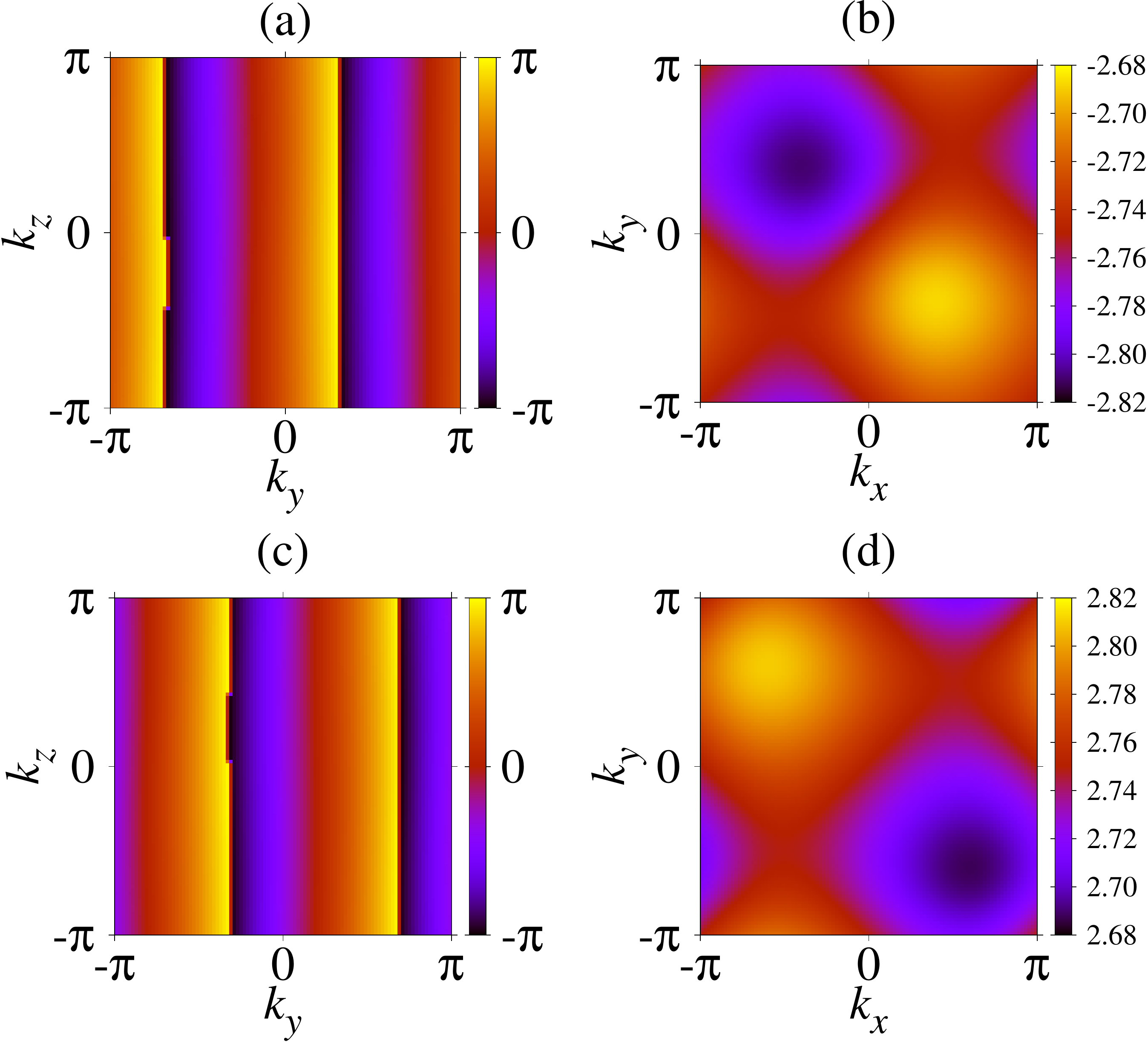}
\end{center}
\caption{\label{Fig_pti_winding} (Color online) Phase of the reflection matrix 
$\Im[\log({\rm det}R_\mu^{(l)})]$  at $\phi=\pm \pi/8$. The lower surface is considered.  The following parameters are assumed: $\beta/(2\pi)=0.23$ and $\delta/(2\pi)=0.4$, which are in the Floquet-topological-insulator phase. The quasienergy is $\phi=\pi/8$ in (a) and (b), and $-\pi/8$ in (c) and (d). 
The surface is normal to the $x$ direction in (a) and (c) and to the $z$ direction in (b) and (d).}
\end{figure}

Figure \ref{Fig_pti_winding} shows $\Im[\log({\rm det}R_\mu^{(l)})]$ in a gapped phase having nontrivial $n_{\mu\nu}^{(l)}$.    
%%%% Fig. 4 %%%%%%%%%%%%%%%%%%%%%%%%%%%%%%
%%%%%%%%%%%%%%%%%%%%%%%%%%%%%%%%%%%%%%%%%%%%%%
In the slab normal to the $x$ direction,  $\Im[\log({\rm det}R_x^{(l)})]$ winds two times in the $k_y$ direction, whereas no winding in the $k_z$ direction.  
On the other hand, in the slab normal to the $z$ direction, no winding is observed for the both $k_x$ and $k_y$ directions. We have observed that the winding is opposite between the upper and lower surfaces by the inversion symmetry, and is common in the gaps around $\phi=\pi/8$ and $-\pi/8$.

We found that for the surface normal to the $z$ direction, $n_{zx}=n_{zy}=0$ regardless of the gaps. In contrast, for the surface normal to the $x$ direction, we can have nonzero $n_{xy}$ of $\pm 2$, though $n_{xz}=0$.  As for the surface normal to the $y$ direction, nonzero $n_{yx}$ can be obtained. In the entire gapped phases, possible combinations of the winding numbers are 
\begin{eqnarray}
&(n_{xy}^{(u)},n_{xz}^{(u)},n_{yz}^{(u)},n_{yx}^{(u)},n_{zx}^{(u)},n_{zy}^{(u)})=(-n,0,0,n,0,0),\label{Eq_winding1}\\ 
&(n_{xy}^{(l)},n_{xz}^{(l)},n_{yz}^{(l)},n_{yx}^{(l)},n_{zx}^{(l)},n_{zy}^{(l)})=(n,0,0,-n,0,0). \label{Eq_winding2}
\end{eqnarray}
In Fig. \ref{Fig_phase}, the index $n$ is also shown as the subscript of symbol ``G''.
The nonzero topological number of $n_{xy}=\pm 2$ implies that there are two surface states whose dispersion relation is extended in the $k_z$ direction. Or in other words, there are two solutions of $k_y$ for the surface states, irrespective of $k_z$.

\begin{figure}
\centerline{
\includegraphics[width=0.23\textwidth]{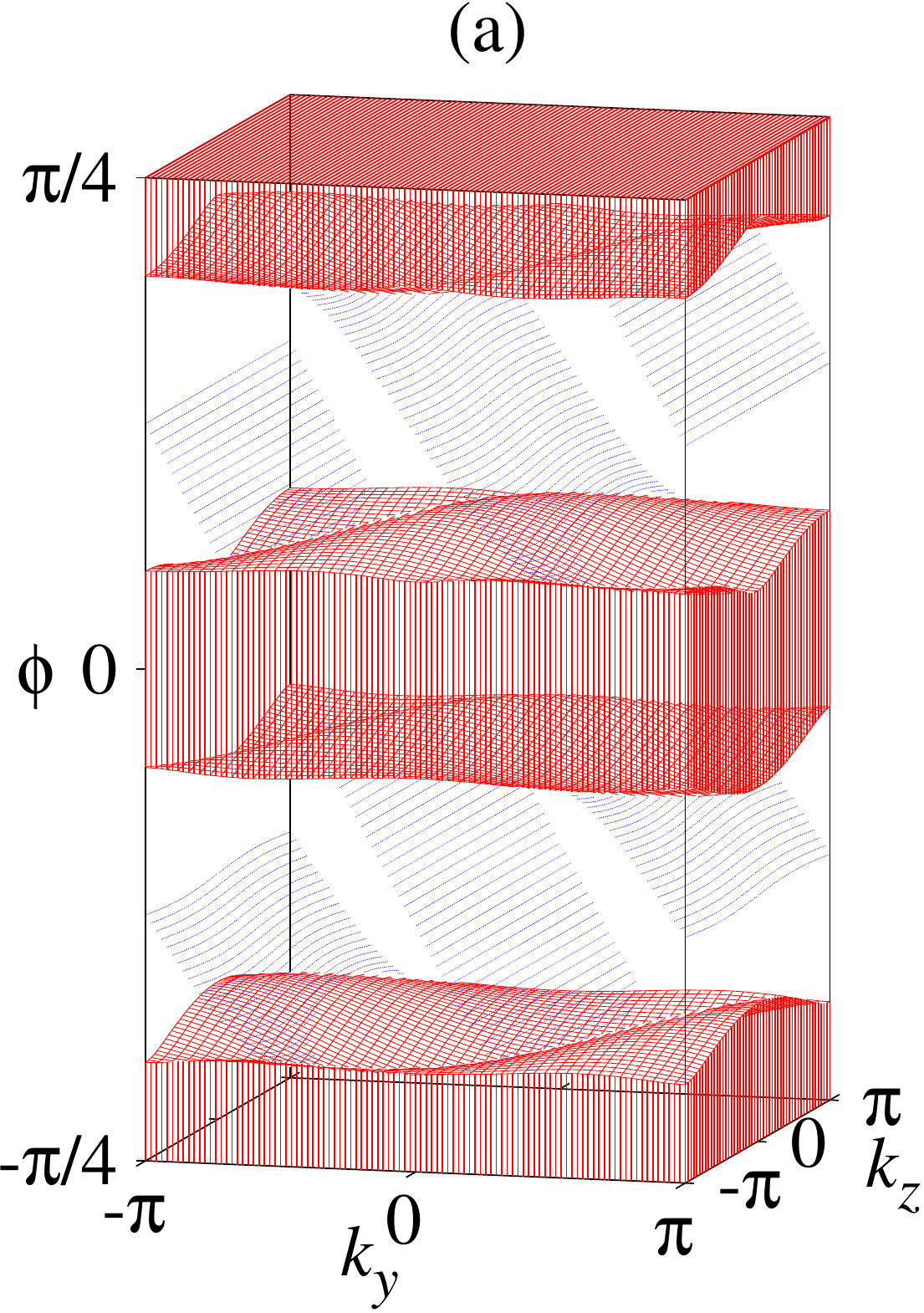} 
\includegraphics[width=0.23\textwidth]{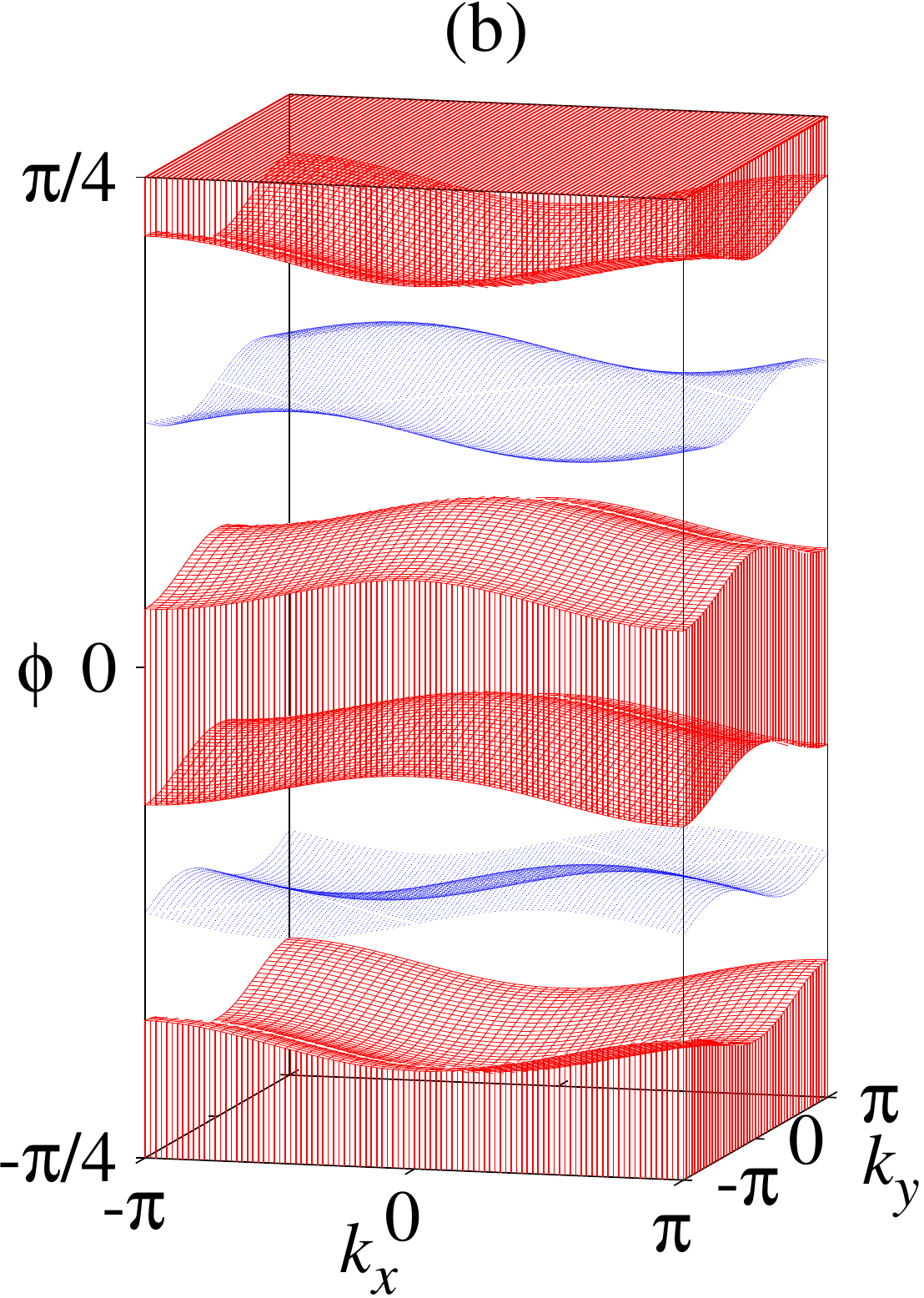} 
}
\caption{\label{Fig_pti_ss} (Color online) Surface states for the lower surfaces normal to the $x$ direction (a) and the $z$ direction (b). The projection of the bulk band structure is also shown. The following parameters are employed: $\beta/(2\pi)=0.23$, $\delta/(2\pi)=0.4$. The  boundary condition is given by Eqs. (\ref{Eq_slabx_bcphase}-\ref{Eq_slabx_bcphase3}) with $p_0=p'_0=q_0=0$ in (a), and by Eq. (\ref{Eq_slabz_bcphase}) with $\varphi_0=0$ in (b). }
\end{figure}

Figure \ref{Fig_pti_ss} shows the surface states in the gapped phase with $n=2$, for the surfaces normal to the $x$ and $z$ directions. The lower surface is considered. 
%%%% Fig. 5 %%%%%%%%%%%%%%%%%%%%%%%%%%%
%%%%%%%%%%%%%%%%%%%%%%%%%%%%%%%%%%%%%%%%
Certainly, for the surface normal to the $x$ direction, we have two surface states (taking account of the periodicity in $k_y$) for each band gap. They are gapless, connecting the upper and lower projected bulk bands. These surface states are reminiscent of the chiral edge state of the 2D ring network lattice \cite{PhysRevLett.110.203904}. The number of two corresponds the two layers per surface unit cell. 
In contrast, the surface normal to the $z$ direction does not hold gapless surface states. The surface states are fully gapped.

\begin{figure}
\begin{center}
\includegraphics[width=0.5\textwidth]{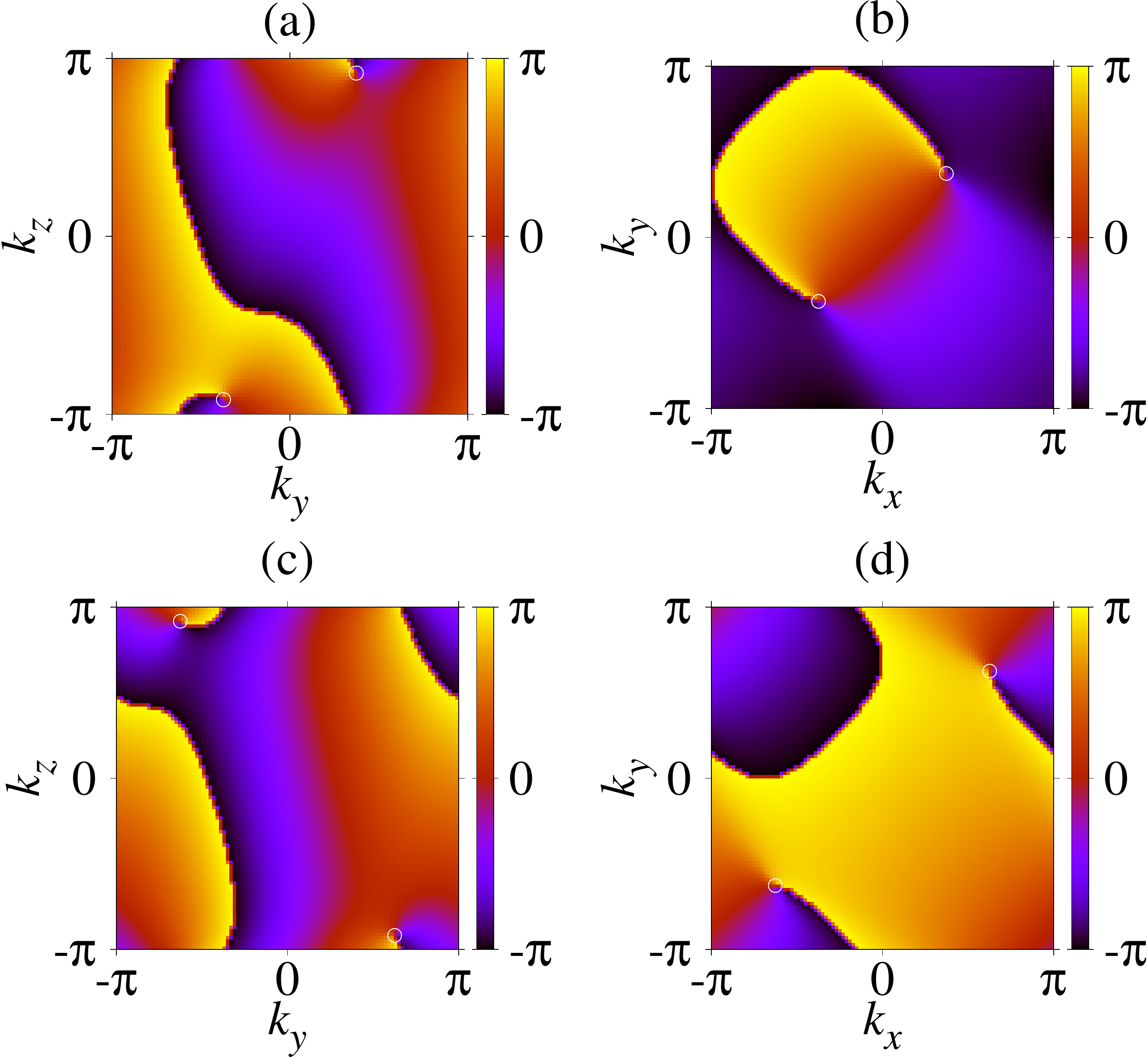}
\end{center}
\caption{\label{Fig_weyl_winding} (Color online) Phase of the reflection matrix 
$\Im[\log({\rm det}R_\mu^{(l)})]$  at $\phi=\pm \pi/8$. The lower surface is considered.  The following parameters are assumed: $\beta/(2\pi)=0.15$ and $\delta/(2\pi)=0.31$, which are in the Weyl phase.  The surface is normal to the $x$ direction in (a) and (c) and to the $z$ direction in (b) and (d). The quasienergy is $\phi=\pi/8$ in (a) and (b), and $-\pi/8$ in (c) and (d). Circles represents the positions of the Weyl points projected onto the surface Brillouin zone.}
\end{figure}

Although the winding number is ill-defined in the Weyl phase, we found that the Weyl phase has a close resemblance to the topological phase with winding number $n=1$, which is absent in our system. As a typical example, we show in Fig. \ref{Fig_weyl_winding}, $\Im[\log({\rm det}R_\mu^{(l)})]$ at the Weyl point quasienergy $\phi=\pm \pi/8$.
%%%% Fig. 6 %%%%%%%%%%%%%%%%%%%%%%%%%%%%%%%%
%%%%%%%%%%%%%%%%%%%%%%%%%%%%%%%%%%%%%%%%%%%%%%%%
We can see the contour maps have two singularities of the Weyl points at $\pm {\bm k}_c$, where $R_\mu^{(l)}$ is no longer unitary and thus $\log({\rm det}R_\mu^{(l)})$ has the real part.  For the surface normal to the $x$ direction, we can observe a winding feature in the $y$ direction. Namely, $\Im[\log({\rm det}R_x^{(l)})]$ winds $2\pi$ if we traverse $k_y$ from $-\pi$ to $\pi$. The winding is not well defined for large $|k_z|$, where the singularity of the Weyl points exists. 
This implies that there is a gapless surface state analogous to those in the FTI phase, in addition to the ordinary Fermi-arc surface state.  
In contrast, $\Im[\log({\rm det}R_z^{(l)})]$  has no winding feature. 
These properties make the surface band structure a bit complicated to visualize.

\begin{figure}
\centerline{
\includegraphics[width=0.25\textwidth]{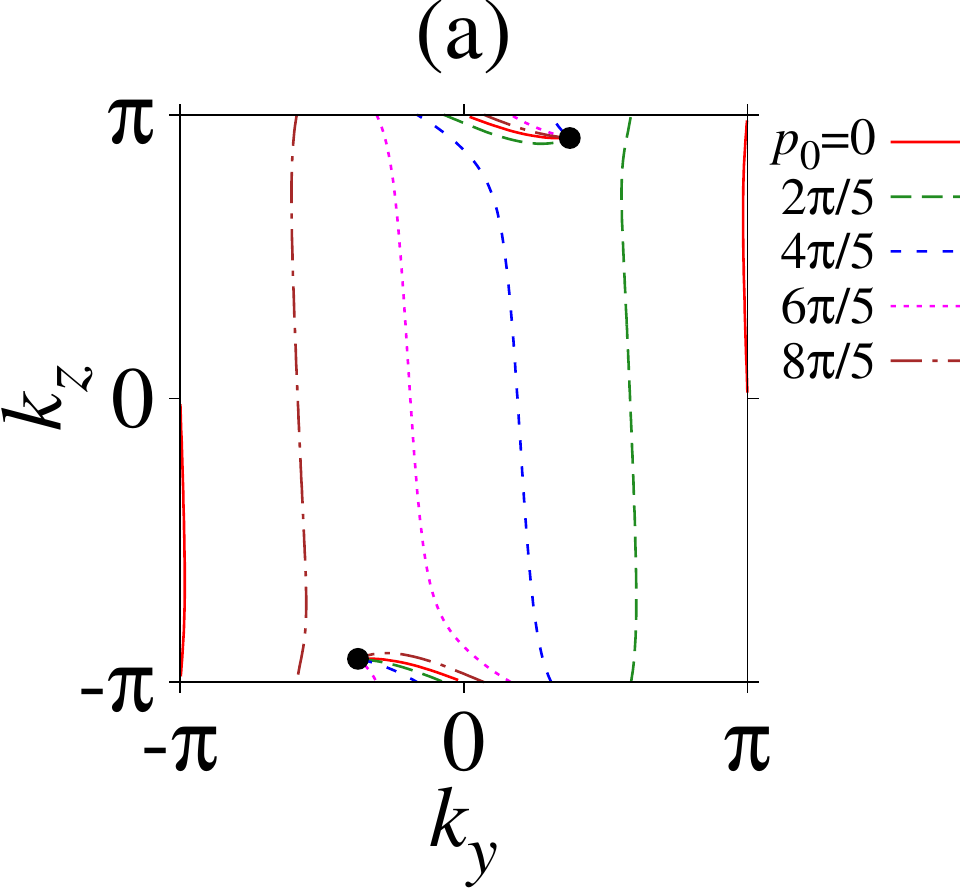}
\includegraphics[width=0.25\textwidth]{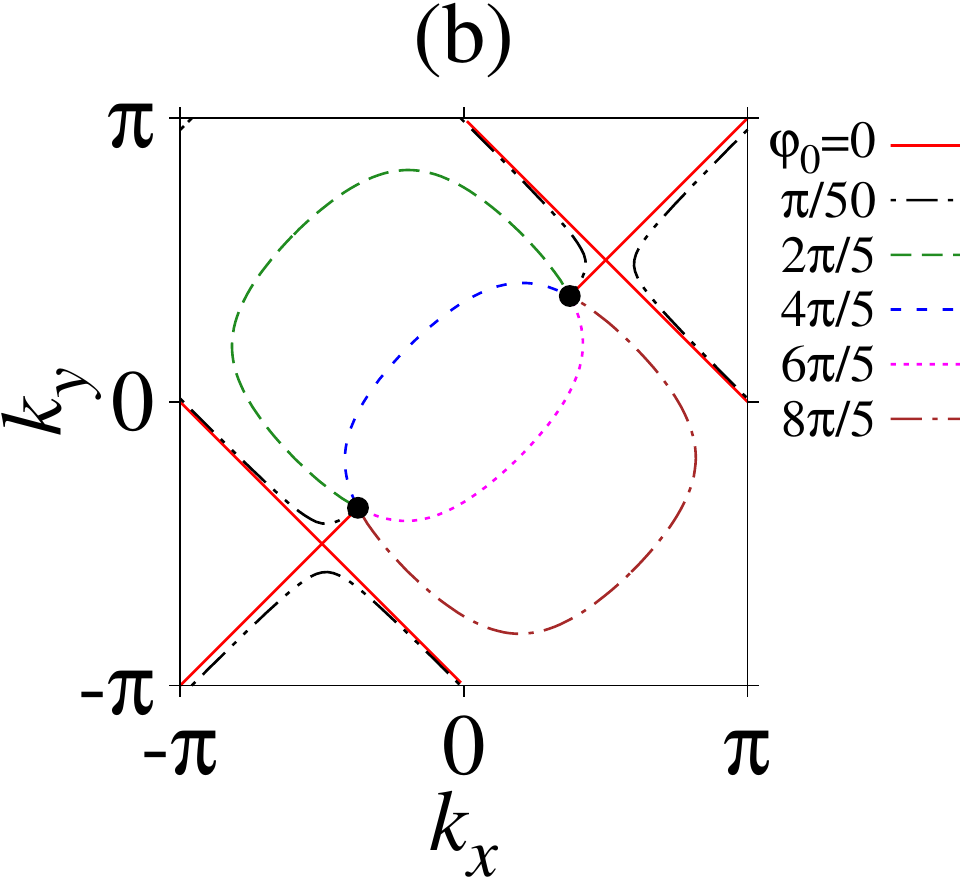}
}
\caption{\label{Fig_arc} (Color online) Equi-quasienergy curves of the Fermi-arc surface states at the Weyl-point energy $\phi=\pi/8$.  
The surface is normal to the $x$ direction (a) and the $z$ direction (b). 
The Weyl points are denoted by the filled circle. 
In (a), the additional phase $p_0$ at the boundary is  changed. The other phases $p'_0$ and  $q_0$ are zero. In (b), the additional phase $\varphi_0$ is changed.  }
\end{figure}

Figure \ref{Fig_arc} shows the equi-quasienergy curves of the surface states at the Weyl-point energy $\phi=\pi/8$. 
%%%% Fig. 7 %%%%%%%%%%%%%%%%%%%%%%%%%%%%%%%%%%%%%%%%%%%%%%
%%%%%%%%%%%%%%%%%%%%%%%%%%%%%%%%%%%%%%%%%%%%%%%%%%%%%%%%%
Here, we introduced nonzero additional phases given in Eqs. (\ref{Eq_slabz_bcphase}) and (\ref{Eq_slabx_bcphase}) to analyze the robustness of the surface states.  
In Fig. \ref{Fig_arc} (a), there are two types of surface states for $p_0=0$, $2\pi/5$, and $8\pi/5$.  One is the standard Fermi arc which connects the two Weyl points in the surface Brillouin zone. The other is similar to the chiral edge states in the FTI phase, whose dispersion curve is extended in the $k_z$ direction.  It is remarkable that at $p_0=4\pi/5$ and $6\pi/5$, the latter surface-state curve merges with the former one, keeping the Fermi-arc feature (connecting the two Weyl points) unchanged.  
Such chiral surface states are absent in the surface normal to the $z$ direction. We note that at $\varphi_0=0$, the Fermi arc has an odd shape, having two cross points. However, they are nothing but a touching of two ``L''-shaped segments of the single Fermi-arc curve. By introducing a small $\varphi_0$, we can see a separation of the curve at the cross points.

The coexistence of the Fermi-arc and chiral surface states are inherent in our system of the stacked 2D ring network lattice.

\section{Optical realization and synthetic gauge fields} 

Let us discuss how to realize our network model optically. 

In our model, a key ingredient is the directional coupling among the rings that separates the counter-clockwise flow from the clockwise one in the rings. To realize this coupling, we introduce additional rings in between the adjacent rings shown in Fig. \ref{Fig_model}. Such a scheme was assumed in the 2D optical ring network in silicon photonics \cite{hafezi2013imaging}. There, the Lieb lattice of the ring resonators are employed. 

With this separation, we can effectively break the time-reversal symmetry in the system. Obviously, the clockwise flow and counter-clockwise flow are time-reversal partners, so that the whole system including two flows is time-reversal symmetric. However, if the separation takes place, we can consider solely the counter-clockwise flow as we did in the paper. There, the time-reversal symmetry appears to be broken. The broken time-reversal symmetry is crucial in the formation of the FW and FTI phases.

\begin{figure}
\centerline{
\includegraphics[width=0.45\textwidth]{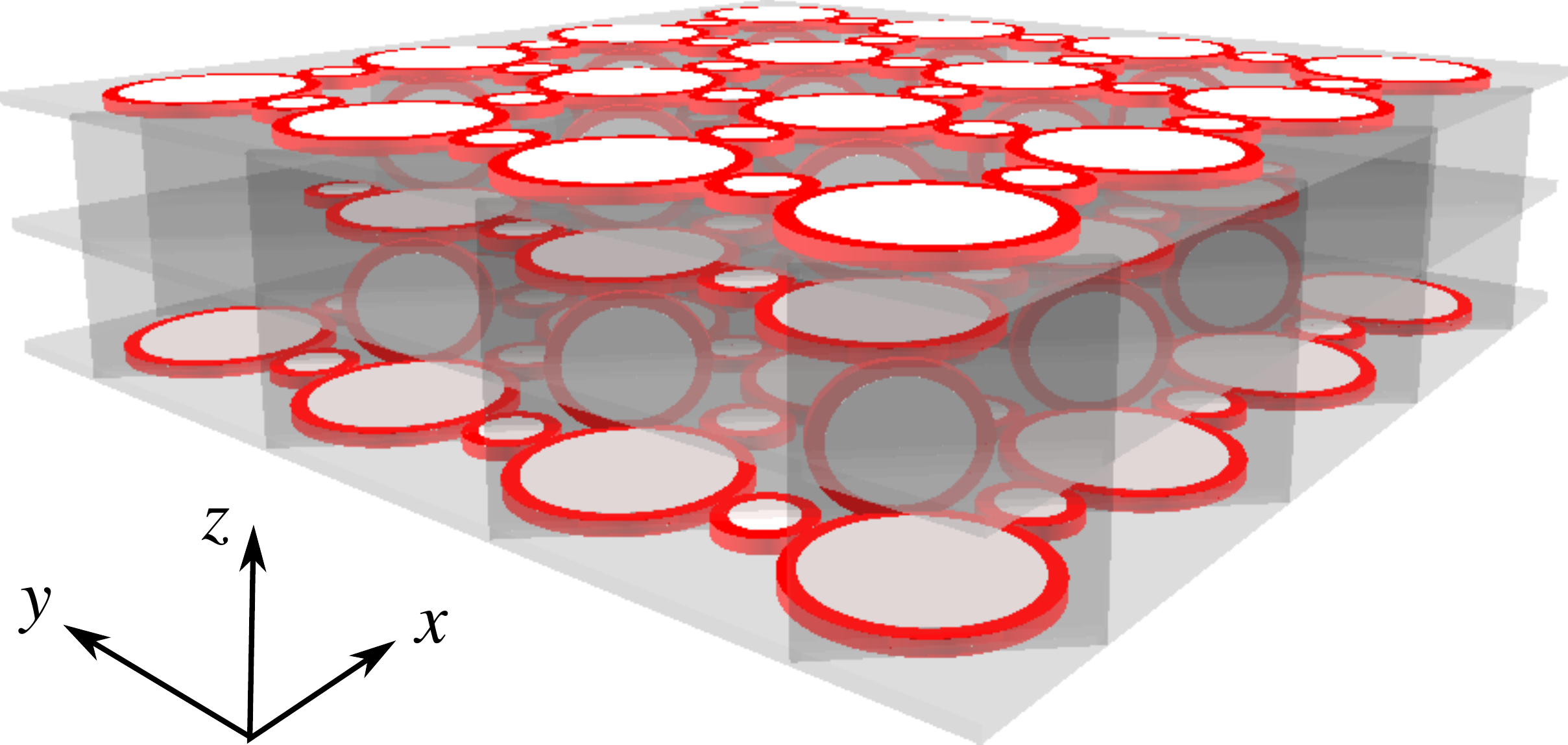}
}
\caption{\label{Fig_optical_realize} (Color online) An optical realization of the stacked 2D ring network model.  Low-index boards that contain the Lieb lattice of high-index ring resonators are stacked periodically in the $z$ direction. A parallel arrangement of the boards that contain the rectangular lattice of the ring resonators are also introduced in the [110] direction.     }
\end{figure}

A proposed optical realization of the stacked 2D ring network is shown in Fig. \ref{Fig_optical_realize}. 
%%%% Fig. 8 %%%%%%%%%%%%%%%%%%%%%%%%%%%%%%%%%
%%%%%%%%%%%%%%%%%%%%%%%%%%%%%%%%%%%%%%%%%%%%%%%%%%%%%%%%
It consists of low-index  boards stacked periodically in the $z$ direction with period 1/2. The Lieb lattice of the high-index ring resonators  is embedded in the board layers. The in-plane shift of (1/2,1/2) is introduced between two adjacent layers. 
In addition, the board layers are  placed periodically with period $1/\sqrt{2}$ along the [110] direction. In these layers, the ring resonators arranged in the rectangular lattice are introduced. Its period is $\sqrt{2}$ in the $[1\bar{1}0]$ direction and 1 in the $z$ direction. The in-plane shift of $1/\sqrt{2}$ in the $[1\bar{1}0]$ direction  and vertical shift of 1/2 are introduced between two adjacent layers. The rings there act as the scattering ports between two adjacent layers in the $z$ direction.

The directional coupling that preserves solely the counter-clockwise flow can be realized as follows. Suppose that two ring resonators are in contact with each other.  If the relevant wavelengths of the ring resonators are much shorter than the contact regions between the two rings, overlapping integrals vanish between the modes with opposite flows near the contact region. Therefore, if the rings in Fig. \ref{Fig_model} are directly contact with each other, two adjacent rings exhibit opposite flows, one for the clockwise flow and the other for the counterclockwise flow. Instead, if we introduce additional rings as in Fig. \ref{Fig_optical_realize}, the counter-clockwise flow in the rings depicted in Fig. \ref{Fig_model} is decoupled from the clockwise one.

\begin{figure}
\centerline{
\includegraphics[width=0.3\textwidth]{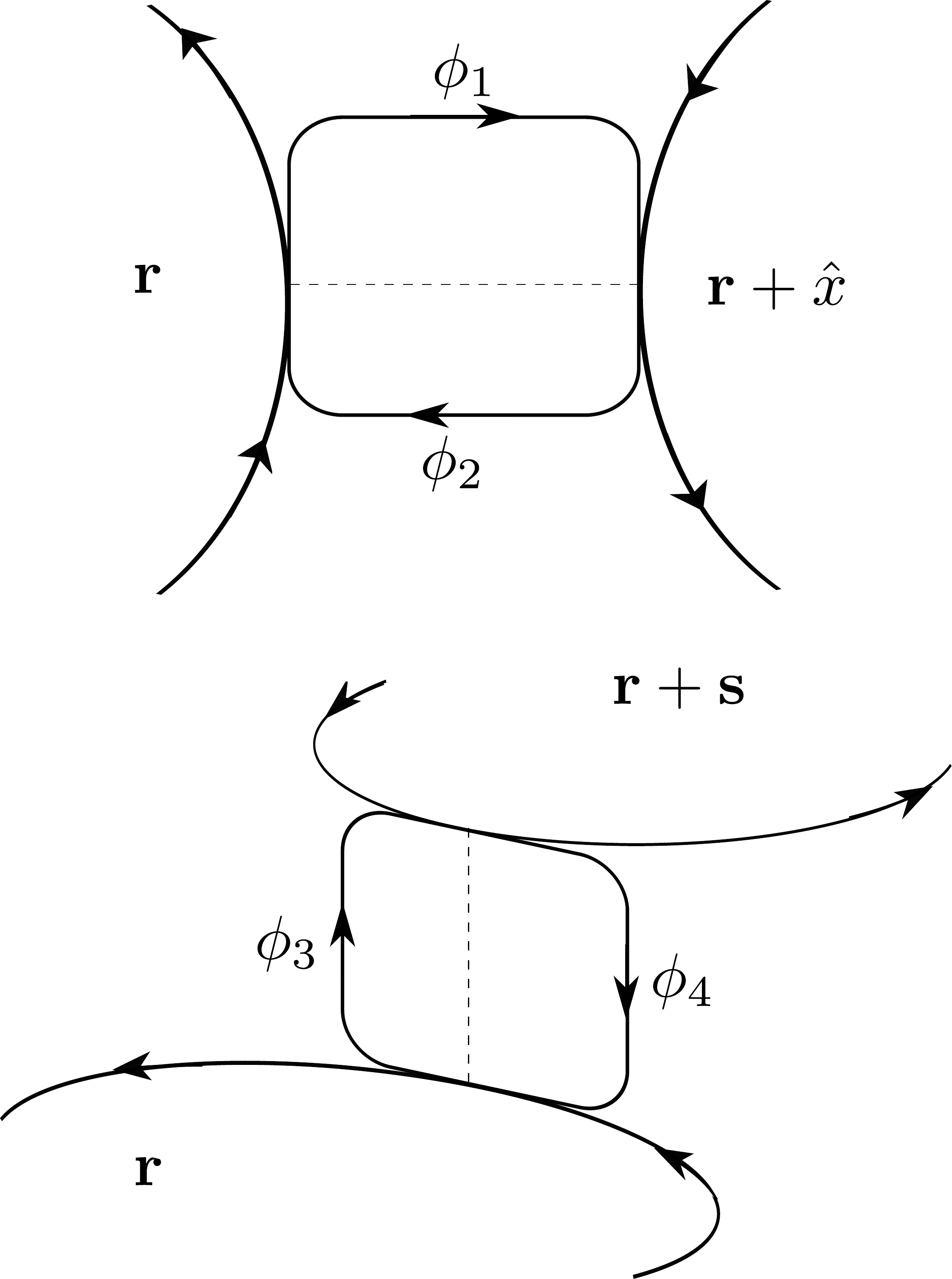}
}
\caption{\label{Fig_gauge} Synthetic gauge field can be implemented by shifting the relative position of the additional ring.  The gauge field is given by the half of the difference in the propagation phases, $\phi_1$ and $\phi_2$ in the $x$ direction and $\phi_3$ and $\phi_4$ in the $z$ direction, of the additional ring.   }
\end{figure}

With this construction, a 3D synthetic gauge field can be implemented just by shifting the relative position of the additional rings, without changing a local structure near the contact points. A schematic illustration is shown in Fig. \ref{Fig_gauge}. 
%%% Fig. 9 %%%%%%%%%%%%%%%%%%%%%%%%%%%
%%%%%%%%%%%%%%%%%%%%%%%%%%%%%%%%%%%%%%%%%%%%%%%%%%
This is a straight-forward extension of the 2D synthetic gauge field proposed by Hafezi et al \cite{hafezi2013imaging}. 
This prescription results in the additional phases to the off-diagonal elements of the hopping S-matrices 
\begin{eqnarray}
&S_i(A) = \left(\begin{array}{cc}
S_i^{++} & S_i^{+-}{\rm e}^{-{\rm i}A_{{\bm r};i}}\\
S_i^{-+}{\rm e}^{{\rm i}A_{{\bm r};i}}& S_i^{--}     
\end{array}\right). 
\end{eqnarray}
Here, $S_i^{\pm\pm}$ are those without the shift, and 
$A_{{\bm r};i}$ is the half of the difference in the propagation phase; $A_{{\bm r};1}=(\phi_1-\phi_2)/2$ and $A_{{\bm r};3}=(\phi_3-\phi_4)/2$, where  $\phi_i$'s are the propagation phase in the additional rings depicted in Fig. \ref{Fig_gauge}. This $A_{{\bm r};i}$ acts as a synthetic gauge field for photons.

In fact, the ring-network system has the gauge invariance under 
\begin{eqnarray}
&\alpha_{\bm r} \to {\rm e}^{{\rm i}\theta_{\bm r}}\alpha_{\bm r},\\
&A_{{\bm r};i} \to A_{{\bm r};i} + \theta_{{\bm r}+{\bm a}_i}-\theta_{\bm r},\\
&{\bm a}_1=\hat{x},\quad {\bm a}_2=\hat{y},\quad {\bm a}_3={\bm s}.   
\end{eqnarray}
Therefore, photons in the ring network system can experience 3D synthetic gauge fields. 
For instance, a synthetic magnetic field can be implemented by a Landau-gauge potential of $A_{{\bm r};i}$. This setting enables us to study 3D Hofstadter butterfly \cite{PhysRevB.45.13488} in a photonic platform. Moreover, various interesting phenomena such as chiral anomaly \cite{PhysRev.177.2426,bell1969pcac} and  chiral magnetic effect \cite{fukushima2008chiral} relevant to Weyl semimetals \cite{PhysRevB.86.115133} may be investigated in photonics.

Finally, let us briefly mention the approximation and limitation in describing the optical ring-resonator system of Fig. \ref{Fig_optical_realize} by  the network model. We must consider a narrow frequency range such that solely a single cavity mode in the ring resonator is relevant. Neglecting other cavity modes than the cavity mode concerned is an important approximation. Another approximation is the diffraction-free propagation of light within the high-index ring-resonator network, resulting in the unitarity of the hopping S-matrices.  In reality, the cavity mode is scattered, to some extent, into free space and the low-index boards. These approximation and limitation should be kept in mind when we adapt the network model.

A detailed numerical analysis of the optical realization and the synthetic gauge field is beyond the scope of the present paper.

\section{Conclusion}

In this paper, we have presented a stacked 2D ring-network lattice as a novel route to 3D FW and FTI phases. They both possess inherent surface states of Fermi arc and gapless types, respectively, in a robust manner. The robustness is guaranteed by the winding number of the reflection matrix in the (pseudo) gap in the bulk. An optical realization of the stacked 2D network model is proposed. It consists of high-index ring resonators embedded in low-index boards. A synthetic gauge field can be implemented rather easily in this realization. Therefore, it is possible to explore effects of a synthetic gauge field in the FW and FTI phase, in a bosonic platform.

%\subsection{}
%\subsubsection{}

% \begin{figure}
% \includegraphics{}%
% \caption{\label{}}
% \end{figure}

% \begin{table}%[H] add [H] placement to break table across pages
% \caption{\label{}}
% \begin{ruledtabular}
% \begin{tabular}{}
% Lines of table here ending with \\
% \end{tabular}
% \end{ruledtabular}
% \end{table}

\appendix
\section{}
In this appendix, we summarize the S-matrix of $N=1$ for the slab in the $x$ direction. 
The derivation of the slab S-matrix is straightforward. 
The basic ingredients are the S-matrices  among the adjacent rings:
\begin{eqnarray}
&\left(\begin{array}{c}
a_0\\
c_1
\end{array}\right)=S_1 \left(\begin{array}{c}
d_0\\
b_1
\end{array}\right){\rm e}^{2{\rm i}\phi},\\
&\left(\begin{array}{c}
b_0\\
d_0
\end{array}\right)=\tilde{S}_2 \left(\begin{array}{c}
e_0\\
f_0
\end{array}\right){\rm e}^{{\rm i}\phi},\\
&\left(\begin{array}{c}
e_0\\
f'_0
\end{array}\right)=S_3 \left(\begin{array}{c}
a_0\\
c'_0
\end{array}\right){\rm e}^{{\rm i}\phi},\\
&\left(\begin{array}{c}
a'_0\\
c'_1
\end{array}\right)=S_1 \left(\begin{array}{c}
d'_0\\
b'_1
\end{array}\right){\rm e}^{2{\rm i}\phi},\\
&\left(\begin{array}{c}
b'_0\\
d'_0
\end{array}\right)=\tilde{S}_2 \left(\begin{array}{c}
e'_0\\
f'_0
\end{array}\right){\rm e}^{{\rm i}\phi},\\
&\left(\begin{array}{c}
e'_0\\
f_1
\end{array}\right)=\tilde{\tilde{S}}_3 \left(\begin{array}{c}
a'_0\\
c_1
\end{array}\right){\rm e}^{{\rm i}\phi},
\end{eqnarray}
where we define 
\begin{eqnarray}
\tilde{\tilde{S}}_3=\left(\begin{array}{cc}
1 & 0\\
0 & {\rm e}^{-{\rm i}(k_y+k_z)}
\end{array}\right)S_3\left(\begin{array}{cc}
1 & 0\\
0 & {\rm e}^{{\rm i}(k_y+k_z)}
\end{array}\right). 
\end{eqnarray}
By removing the amplitudes absent in Eq. (\ref{Eq_slabx}) with $N=1$, we obtain 
\begin{eqnarray}
&S_{x;1}(1,1)={\rm e}^{3{\rm i}\phi}(\tilde{\tilde{S}}_3^{-+}A'_c + \tilde{\tilde{S}}_3^{--})S_1^{-+}D_f \nonumber \\
&\hskip40pt +{\rm e}^{2{\rm i}\phi}\tilde{\tilde{S}}_3^{-+}A'_f S_3^{-+}D_f,\\
&S_{x;1}(1,2)={\rm e}^{3{\rm i}\phi}(\tilde{\tilde{S}}_3^{-+}A'_c + \tilde{\tilde{S}}_3^{--})S_1^{-+}D_c \nonumber \\
&\hskip40pt +{\rm e}^{2{\rm i}\phi}\tilde{\tilde{S}}_3^{-+}A'_f (S_3^{-+}D_c + S_3^{--}),\\ 
&S_{x;1}(1,3)={\rm e}^{3{\rm i}\phi}(\tilde{\tilde{S}}_3^{-+}A'_c + \tilde{\tilde{S}}_3^{--})(S_1^{-+}D_b+S_1^{--}) \nonumber \\
&\hskip40pt +{\rm e}^{2{\rm i}\phi}\tilde{\tilde{S}}_3^{-+}A'_f S_3^{-+}A_b,\\
&S_{x;1}(1,4)={\rm e}^{{\rm i}\phi}\tilde{\tilde{S}}_3^{-+}A'_b,\\
&S_{x;1}(2,1)={\rm e}^{4{\rm i}\phi}S_1^{-+}D'_c S_1^{-+}D_f \nonumber \\
&\hskip40pt + {\rm e}^{3{\rm i}\phi}S_1^{-+}D'_f S_3^{-+}A_f,\\
&S_{x;1}(2,2)={\rm e}^{4{\rm i}\phi}S_1^{-+}D'_c S_1^{-+}D_c \nonumber \\
&\hskip40pt + {\rm e}^{3{\rm i}\phi}S_1^{-+}D'_f (S_3^{-+}A_c + S_3^{--}),\\
&S_{x;1}(2,3)={\rm e}^{4{\rm i}\phi}S_1^{-+}D'_c (S_1^{-+}D_b+ S_1^{--}) \nonumber \\
&\hskip40pt + {\rm e}^{3{\rm i}\phi}S_1^{-+}D'_f S_3^{-+}A_b,\\
&S_{x;1}(2,4)={\rm e}^{2{\rm i}\phi}(S_1^{-+}D'_b +S_1^{--}),\\
&S_{x;1}(3,1)={\rm e}^{{\rm i}\phi}(\tilde{S}_2^{++}E_f + \tilde{S}_2^{+-}),\\
&S_{x;1}(3,2)={\rm e}^{{\rm i}\phi}\tilde{S}_2^{++}E_c,\\
&S_{x;1}(3,3)={\rm e}^{{\rm i}\phi}\tilde{S}_2^{++}E_b,\\
&S_{x;1}(3,4)=0,\\
&S_{x;1}(4,1)={\rm e}^{3{\rm i}\phi}\tilde{S}_2^{++}E'_cS_1^{-+}D_f \nonumber \\
&\hskip40pt +{\rm e}^{2{\rm i}\phi}(\tilde{S}_2^{++}E'_f + \tilde{S}_2^{+-}) S_3^{-+}A_f,\\
&S_{x;1}(4,2)={\rm e}^{3{\rm i}\phi}\tilde{S}_2^{++}E'_cS_1^{-+}D_c \nonumber \\
&\hskip40pt +{\rm e}^{2{\rm i}\phi}(\tilde{S}_2^{++}E'_f + \tilde{S}_2^{+-}) (S_3^{-+}A_c+S_3^{--}),\\
&S_{x;1}(4,3)={\rm e}^{3{\rm i}\phi}\tilde{S}_2^{++}E'_c(S_1^{-+}D_b + S_1^{--}) \nonumber \\
&\hskip40pt +{\rm e}^{2{\rm i}\phi}(\tilde{S}_2^{++}E'_f + \tilde{S}_2^{+-}) S_3^{-+}A_b,\\
&S_{x;1}(4,4)={\rm e}^{{\rm i}\phi}\tilde{S}_2^{++}E'_b.
\end{eqnarray}
Here, the coefficients $A_b$ etc are given by 
\begin{eqnarray}
&A_b=(1-{\rm e}^{4{\rm i}\phi}S_1^{++}\tilde{S}_2^{-+}S_3^{++})^{-1}{\rm e}^{2{\rm i}\phi}S_1^{+-},\\
&A_c=(1-{\rm e}^{4{\rm i}\phi}S_1^{++}\tilde{S}_2^{-+}S_3^{++})^{-1}{\rm e}^{4{\rm i}\phi}S_1^{++}\tilde{S}_2^{-+}S_3^{+-},\\
&A_f=(1-{\rm e}^{4{\rm i}\phi}S_1^{++}\tilde{S}_2^{-+}S_3^{++})^{-1}{\rm e}^{3{\rm i}\phi}S_1^{++}\tilde{S}_2^{--},\\
&D_b=(1-{\rm e}^{4{\rm i}\phi}S_1^{++}\tilde{S}_2^{-+}S_3^{++})^{-1}{\rm e}^{4{\rm i}\phi}S_1^{+-}\tilde{S}_2^{-+}S_3^{++},\\
&D_c=(1-{\rm e}^{4{\rm i}\phi}S_1^{++}\tilde{S}_2^{-+}S_3^{++})^{-1}{\rm e}^{2{\rm i}\phi}\tilde{S}_2^{-+}S_3^{+-},\\
&D_f=(1-{\rm e}^{4{\rm i}\phi}S_1^{++}\tilde{S}_2^{-+}S_3^{++})^{-1}{\rm e}^{{\rm i}\phi}\tilde{S}_2^{--},\\
&E_b=(1-{\rm e}^{4{\rm i}\phi}S_1^{++}\tilde{S}_2^{-+}S_3^{++})^{-1}{\rm e}^{3{\rm i}\phi}S_1^{+-}S_3^{++},\\
&E_c=(1-{\rm e}^{4{\rm i}\phi}S_1^{++}\tilde{S}_2^{-+}S_3^{++})^{-1}{\rm e}^{{\rm i}\phi}S_3^{+-},\\
&E_f=(1-{\rm e}^{4{\rm i}\phi}S_1^{++}\tilde{S}_2^{-+}S_3^{++})^{-1}{\rm e}^{4{\rm i}\phi}S_1^{+-}\tilde{S}_2^{--}S_3^{++},\\
&A'_b=(1-{\rm e}^{4{\rm i}\phi}S_1^{++}\tilde{S}_2^{-+}\tilde{\tilde{S}}_3^{++})^{-1}{\rm e}^{2{\rm i}\phi}S_1^{+-},\\
&A'_c=(1-{\rm e}^{4{\rm i}\phi}S_1^{++}\tilde{S}_2^{-+}\tilde{\tilde{S}}_3^{++})^{-1}{\rm e}^{4{\rm i}\phi}S_1^{++}\tilde{S}_2^{-+}\tilde{\tilde{S}}_3^{+-},\\
&A'_f=(1-{\rm e}^{4{\rm i}\phi}S_1^{++}\tilde{S}_2^{-+}\tilde{\tilde{S}}_3^{++})^{-1}{\rm e}^{3{\rm i}\phi}S_1^{++}\tilde{S}_2^{--},\\
&D'_b=(1-{\rm e}^{4{\rm i}\phi}S_1^{++}\tilde{S}_2^{-+}\tilde{\tilde{S}}_3^{++})^{-1}{\rm e}^{4{\rm i}\phi}S_1^{+-}\tilde{S}_2^{-+}\tilde{\tilde{S}}_3^{++},\\
&D'_c=(1-{\rm e}^{4{\rm i}\phi}S_1^{++}\tilde{S}_2^{-+}\tilde{\tilde{S}}_3^{++})^{-1}{\rm e}^{2{\rm i}\phi}\tilde{S}_2^{-+}\tilde{\tilde{S}}_3^{+-},\\
&D'_f=(1-{\rm e}^{4{\rm i}\phi}S_1^{++}\tilde{S}_2^{-+}\tilde{\tilde{S}}_3^{++})^{-1}{\rm e}^{{\rm i}\phi}\tilde{S}_2^{--},\\
&E'_b=(1-{\rm e}^{4{\rm i}\phi}S_1^{++}\tilde{S}_2^{-+}\tilde{\tilde{S}}_3^{++})^{-1}{\rm e}^{3{\rm i}\phi}S_1^{+-}\tilde{\tilde{S}}_3^{++},\\
&E'_c=(1-{\rm e}^{4{\rm i}\phi}S_1^{++}\tilde{S}_2^{-+}\tilde{\tilde{S}}_3^{++})^{-1}{\rm e}^{{\rm i}\phi}\tilde{\tilde{S}}_3^{+-},\\
&E'_f=(1-{\rm e}^{4{\rm i}\phi}S_1^{++}\tilde{S}_2^{-+}\tilde{\tilde{S}}_3^{++})^{-1}{\rm e}^{4{\rm i}\phi}S_1^{+-}\tilde{S}_2^{--}\tilde{\tilde{S}}_3^{++}.
\end{eqnarray}

%\begin{acknowledgments}
\ack  %IOP 
This work was partially supported by KAKENHI Grant No. 26390013. 
%\end{acknowledgments}

\section*{References}
%\bibliography{../../Database/mydata}
\providecommand{\newblock}{}

\end{document}